# Room-temperature mid-infrared detection using metasurface-absorber-integrated phononic crystal oscillator

Zichen Xi, Zengyu Cen, Dongyao Wang, Joseph G. Thomas, Bernadeta R. Srijanto, Ivan I. Kravchenko, Jiawei Zuo, Honghu Liu, Jun Ji, Yizheng Zhu, Yu Yao*, and Linbo Shao*

Z. Xi, J. G. Thomas, J. Ji, Y. Zhu, and L. Shao
Bradley Department of Electrical and Computer Engineering
Virginia Tech, Blacksburg, VA 24061, USA

Z. Cen, D. Wang, J. Zuo, and Y. Yao
School of Electrical, Computer and Energy Engineering, Center for Photonic Innovation
Arizona State University, Tempe, AZ, 85281 USA

J. G. Thomas, and Y. Zhu
Center for Photonics Technology
Virginia Tech, Blacksburg, VA 24061, USA

B. R. Srijanto, I. I. Kravchenko
Center for Nanophase Materials Sciences
Oak Ridge National Laboratory, Oak Ridge, TN 37830 USA

H. Liu
Department of Mathematics
Virginia Tech, Blacksburg, VA 24061, USA

L. Shao
Department of Physics and Center for Quantum Information Science and Engineering (VTQ)
Virginia Tech, Blacksburg, VA 24061, USA

Z. Xi, Z. Cen, and D. Wang contributed equally to this work
*Corresponding authors: yuyao@asu.edu; shaolb@vt.edu

## ABSTRACT

Mid-infrared (MIR) detectors find extensive applications in chemical sensing, spectroscopy, communications, biomedical diagnosis and space explorations. Alternative to semiconductor MIR photodiodes and bolometers, mechanical-resonator-based MIR detectors show advantages in higher sensitivity and lower noise at room temperature, especially towards longer wavelength infrared. Here, we demonstrate uncooled room-temperature MIR detectors based on lithium niobate surface acoustic wave phononic crystal (PnC) resonators integrated with wavelength-and-polarization-selective metasurface absorber arrays. The detection is based on the resonant frequency shift induced by the local temperature change due to MIR absorptions. The PnC resonator is configured in an oscillating mode, enabling active readout and low frequency noise. Compared with detectors based on tethered thin-film mechanical resonators, our non-suspended, fully supported PnC resonators offer lower noise, faster thermal response, and robustness in both fabrication and practical applications. Our 1-GHz oscillator-based MIR detector shows a relative frequency deviation of $5.24 \times 10^{-10}$ $\mathrm{Hz}^{-1/2}$ at an integration time of 50 μs, leading to an incident noise equivalent power of 197 pW/$\sqrt{\mathrm{Hz}}$ when input 6-μm MIR light is modulated at 1.8 kHz, and a large dynamic range of $10^7$ in incident MIR power. Our device architecture is compatible with the scalable manufacturing process and can be readily extended to a broader spectral range by tailoring the absorbing wavelengths of metasurface absorbers.

**KEYWORDS:** mid-infrared detection, metasurface absorber, phononic crystal, mechanical oscillator

## 1. INTRODUCTION

Detection of mid-infrared (MIR) radiation, 3 μm to 50 μm wavelength[1], is the cornerstone of MIR applications including astronomy[2], chemical analysis[3-4], biomedicine[5-6], agriculture[7], and communications[8]. A desirable MIR detector would feature high sensitivity, fast response, a large dynamic range, room temperature operation, and a widely designable wavelength range. MIR detectors have been developed based on semiconductor photodiodes, bolometers, nonlinear optics, and mechanical resonators. Semiconductor MIR photodetectors based on mercury cadmium telluride (MCT)[9-10], indium antimonide and aluminum indium arsenide antimonide (InSb/AlInAsSb) material systems[11-13], quantum well structures[14-15], and two-dimensional materials[16-20], have been demonstrated with high responsivity and speed, yet they usually require cryogenic temperatures to achieve high sensitivity due to small photon energy at MIR wavelengths. Moreover, their operation wavelength range is limited by the semiconductor bandgap[9-13] or band offset[14-15], and the detection sensitivity drops significantly for wavelength in the long wave infrared spectral range and beyond (>8 μm). Nonlinear optical approaches[21-27] that up-convert MIR to near infrared or visible can leverage high performance photodiodes or image sensors at shorter wavelengths, although achieving a high nonlinear conversion efficiency is still challenging on chip. Bolometers based on two-dimensional materials[28-29] and semiconductors[30] demonstrate room-temperature MIR detection yet are still limited by high noise level and low sensitivity at room temperature.

Compared to photodetectors and bolometers, mechanical resonator-based mid-infrared (MIR) detectors show advantages in room-temperature MIR detection, especially at wavelengths longer than 5 μm[31-33]. As a temperature-based approach, such mechanical detectors could cover a broad range of wavelengths, either using broadband absorbers, such as platinum thin films[34], or metamaterial optical absorbers[35]. To date, most mechanical-based MIR detectors leverage suspended mechanical resonators with narrow and long supports tethers (for example, 5-μm wide and 600-μm long tethers in Ref. [31]) for high mechanical quality factors and ultimate thermal isolations, resulting in ultrahigh responsivities in relative frequency shift (11,000 $W^{-1}$ Ref. [31]). However, such tethered structures result in relatively slow thermal response time of a few milliseconds and post challenges for scalable manufacturing and applications in harsh environment, such as under high accelerations.

Here, we demonstrate MIR detectors leveraging low-noise surface acoustic wave phononic crystal (PnC) oscillator on bulk lithium niobate (LN) with co-designed wavelength-and-polarization-selective metamaterial absorbers. The oscillation configuration of our detector enables a direct and large-bandwidth readout of the resonant frequency, as well as reduces the frequency noises by over two orders of magnitude compared to a passive measurement configuration. Our oscillator-based MIR detector features an effective area, defined by its acoustic-wave mode area, of 332 $\mu m^2$ and a low relative frequency noise of $5.24 \times 10^{-10}$ $Hz^{-1/2}$ at an integration time of 50 μs at its oscillation frequency of 1 GHz. The metamaterial absorbers integrated on our PnC resonator offer wavelength and polarization selective MIR absorption. We demonstrate an incident noise equivalent power (NEP) of 197 pW/$\sqrt{\text{Hz}}$ when an input 6-μm MIR light is modulated at 1.8 kHz. Our detector features a large dynamic range of $10^7$ in incident MIR power from 100s pW to over 1 mW. By changing the absorber designs, our device architecture could potentially enable monolithic integration of detectors for the full optical spectrum from ultraviolet, to far infrared, and THz waves on a single chip.

## 2. METASURFACE-ABSORBER-INTEGRATED PNC RESONATOR FOR MIR DETECTION

We develop PnC resonator integrated with metasurface absorbers for MIR detection (**Figs. 1(a)-1(c)**). Our PnC resonator is defined by etched grooves with varying periods and widths in different segments[36-37]. Two interdigital transducers (IDTs) – one located outside, and the other located inside the PnC resonator – provide efficient electromechanical coupling to our PnC resonator. Metasurface absorbers based on coupled plasmonic antennas are placed between the etched grooves at the resonant mode area. The incident MIR is transduced into heat by the absorbers, induces local temperature change, and results in the frequency shift of our PnC resonator via the temperature-dependent elasticity and thermal expansion. As proof of concept, we demonstrate one single-wavelength absorber design for 6 μm wavelength (**Figs. 1(d) and 1(e)**), and another dual-wavelength absorber design for 6 μm and 9 μm (**Figs. 1(f) and 1(g)**). Our absorbers are made of a thin metal stack of 10-nm-thick chromium and 30-nm-thick gold. We note that a small absorber area is preferred to minimize the mass loading on the PnC structures and the resistive electric loss to the mechanical mode due to the piezoelectric effect of the lithium niobate substrate.

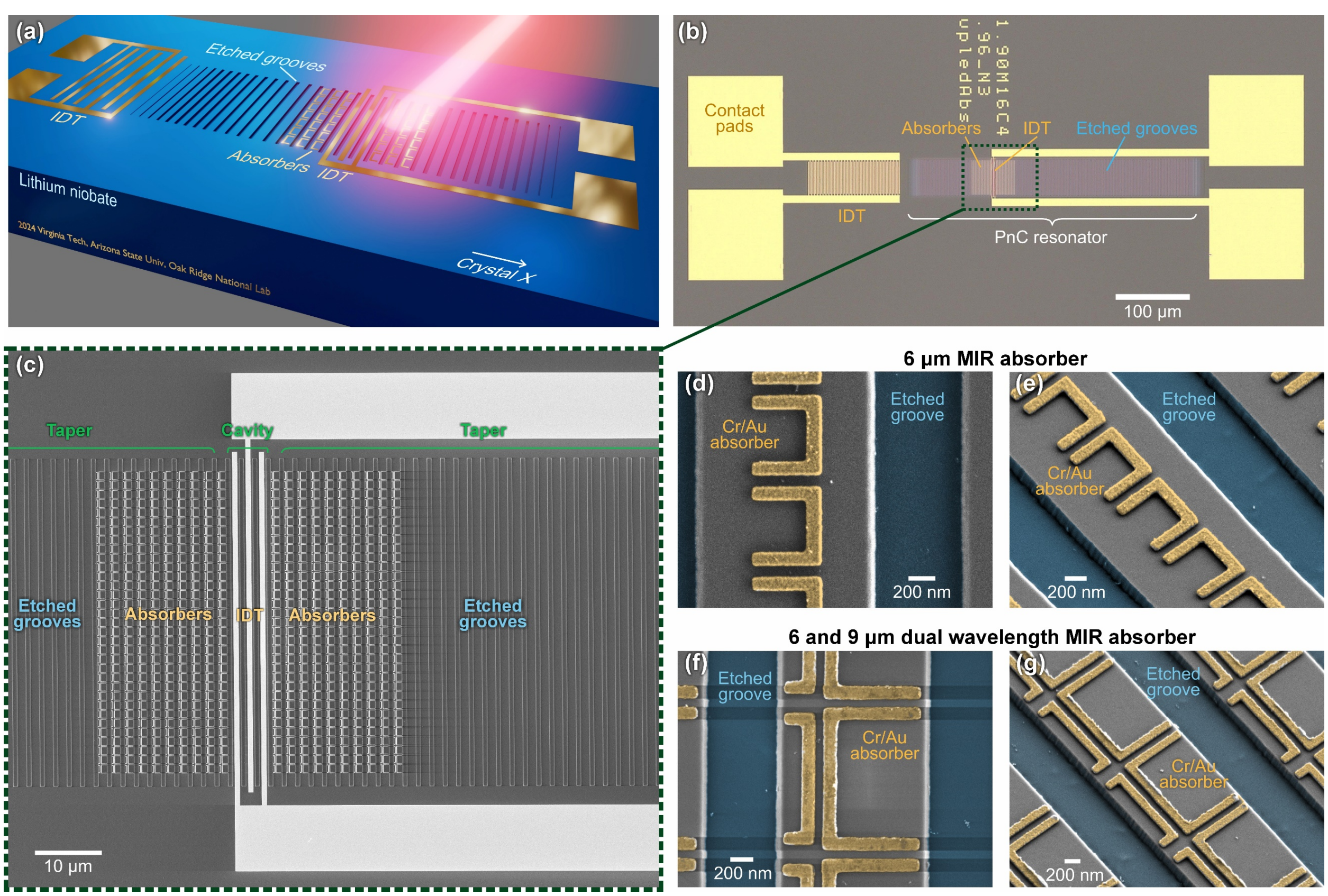

**FIG. 1. Phononic crystal (PnC) mid-infrared (MIR) detector.** (a) Illustration of the detector fabricated on a 128° Y cut lithium niobate (LN) platform, composed of a phononic crystal (PnC) resonator defined by etched grooves, a pair of interdigital transducers (IDT), and an MIR absorber array. IDTs are employed to excite and receive the acoustic wave. The oscillation is formed with sufficient gain provided by a low noise amplifier. The device is fabricated on 128° Y cut lithium niobate with acoustic waves propagating in crystal X direction. (b) Microscope image of fabricated detector. IDT and absorbers are made of 10/30 nm Chromium/gold. (c) Scanning electron microscopy (SEM) image of the PnC resonator with the absorber array. IDT electrodes and absorbers are placed between the etched grooves in the PnC cavity and taper region. False colored SEM images of (d)(e) the 6-μm single-wavelength absorbers and (f)(g) the 6/9-μm dual-wavelength absorbers.

We numerically and experimentally characterize the absorbers (**Fig. 2**). The 6 μm absorbers employ a C-shape design. At the peak absorption wavelength, the simulated profile shows an enhanced optical electric field in the gap between the C-shape arms (**Fig. 2(a)**). At the favorable 0° polarization, the numerical simulation shows a reflection peak of 53% at 6.10 μm and an absorption peak of 30% at 6.17 μm (**Fig. 2(c)**). The experimentally measured reflection spectrum of the 6-μm absorber array shows a peak of 49% at 6.05 μm (**Fig. 2(c)**), agreeing with our simulation. The measured reflectance at different polarization angles shows a contrast of 5 between the favorable 0° polarization and insensitive 90° polarization (**Fig. 2(e)**).

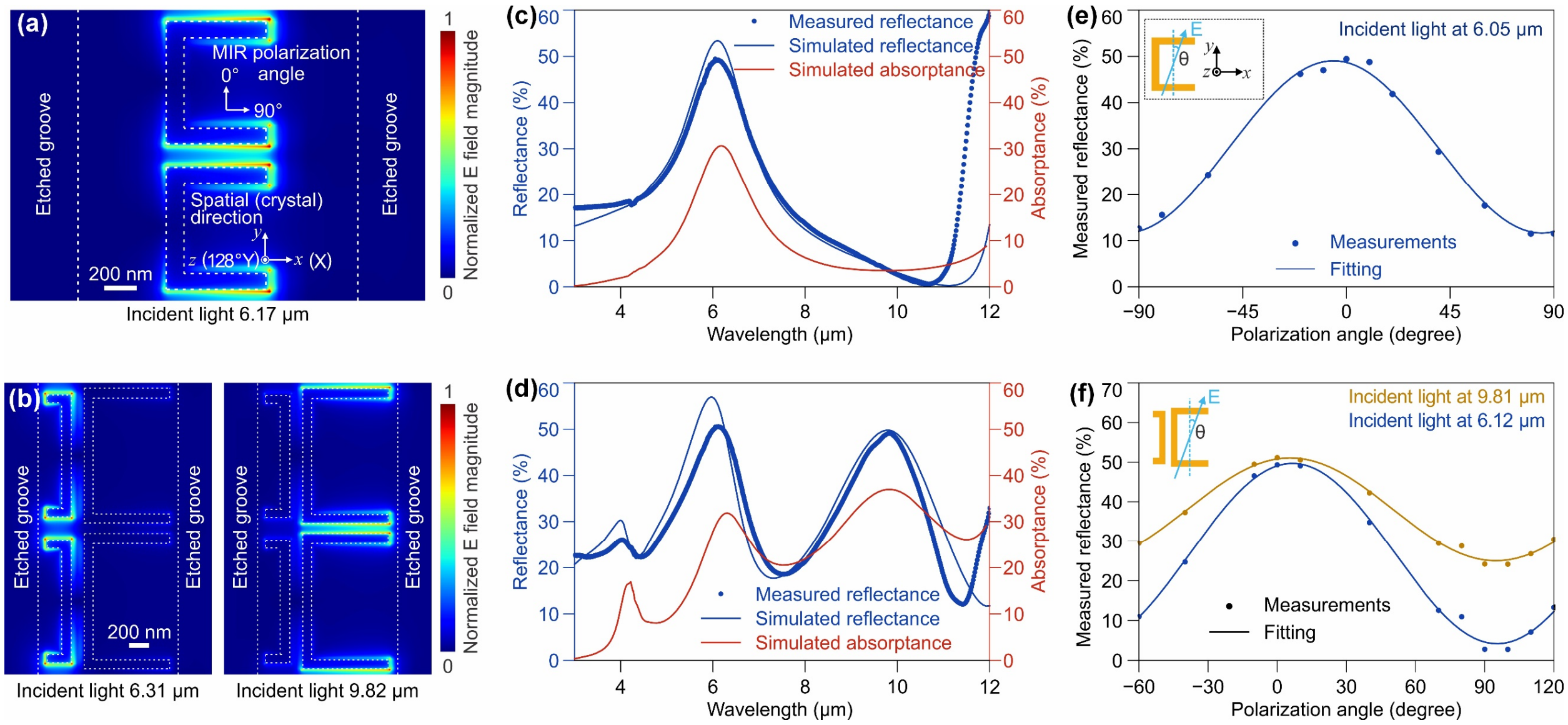

**FIG. 2. Mid-infrared (MIR) absorbers for 6 μm single wavelength and 6-and-9 μm dual wavelengths.** (a)(b) Numerically simulated near field magnitude distribution at the peak absorption wavelengths. The boundary of the metal absorber and phononic crystal etched grooves are marked by the dashed lines. (c)(d) Simulated reflection and absorption spectra and experimentally measured reflection spectra. (e)(f) Experimentally measured polarization-dependent reflection at peak absorption wavelengths. The measured results are fitted to a sine curve. Solid lines represent fitted curves, and dots indicate measured data. The insets illustrate the definition of polarization angle $\theta$, which is the angle between the incident electric field vector $E$ (blue solid arrow) and the antenna axis (blue dashed line). The propagation direction of incident light is normal to the $x$-$y$ plane. (a)(c)(e) are of the 6-μm single-wavelength absorbers, and (b)(d)(f) are of the 6/9-μm dual-wavelength absorbers.

The 6-and-9-μm dual-wavelength absorbers employ a double C-shape design. A 6.31-μm incident MIR is resonant with the left narrower C-shape plasmonic antenna, while a 9.81-μm incident MIR is resonant with the right larger C-shape plasmonic antenna (**Fig. 2(b)**). At the favorable 0° polarization, the simulated reflection spectrum shows peaks of 57% and 50% at 5.97 and 9.77 μm wavelengths, respectively; the simulated absorption spectrum shows peaks of 32% and 37% at 6.31 and 9.82 μm, respectively (**Fig. 2(d)**). The experimentally measured reflection spectrum shows a peak of 51% and 49% at 6.12 and 9.81 μm, respectively, agreeing with our simulation. The measured reflectance at different polarization angles shows a contrast of 10 at 6.12 μm and a lower contrast of 2 at 9.81 μm due to the higher absorption of lithium niobate substrate at this wavelength (**Fig. 2(f)**). Details of our absorber designs and simulations are provided in **Supplemental Note 1**.

## 3. OSCILLATOR BASED ON PNC RESONATOR FOR MIR DETECTION

Our PnC resonators[36-37] are fabricated on a 128°Y-cut LN substrate with acoustic waves propagating along crystal X axis. This 128°Y-X configuration offers a low propagation loss, small divergence, and efficient electromechanical coupling for the acoustic waves. The temperature coefficient for the phase velocity of the 128Y-X acoustic waves is -71.2 ppm/K [37-38]. It is greater than the coefficients of silicon (-60 ppm/K)[33] and silicon nitride (-30 ppm/K)[31], both are widely used in trampoline detectors. We design the PnC resonator by adjusting periods and widths for etched grooves in different segments (**Fig. 3(a)**). The etched grooves with a period of 1.96 μm in the mirror segments gradually increases to 2.00 μm in the cavity segment. Details of the PnC resonator design are provided in **Fig. S1**. The PnC resonator supports a mode at 1 GHz with a loaded $Q$ factor of 1,000, leading to a $fQ$ product of $10^{12}$, and an electrical transmission (IDT1 to IDT2) of -28 dB (**Figs. S2(b) and S2(c)**). We measure the acoustic mode profile (**Figs. 3(b) and 3(c)**) using our home-built optical vibrometer[39]. The acoustic resonant mode shows a mode area of 332 $\mu m^2$ (**Supplemental Note 2**).

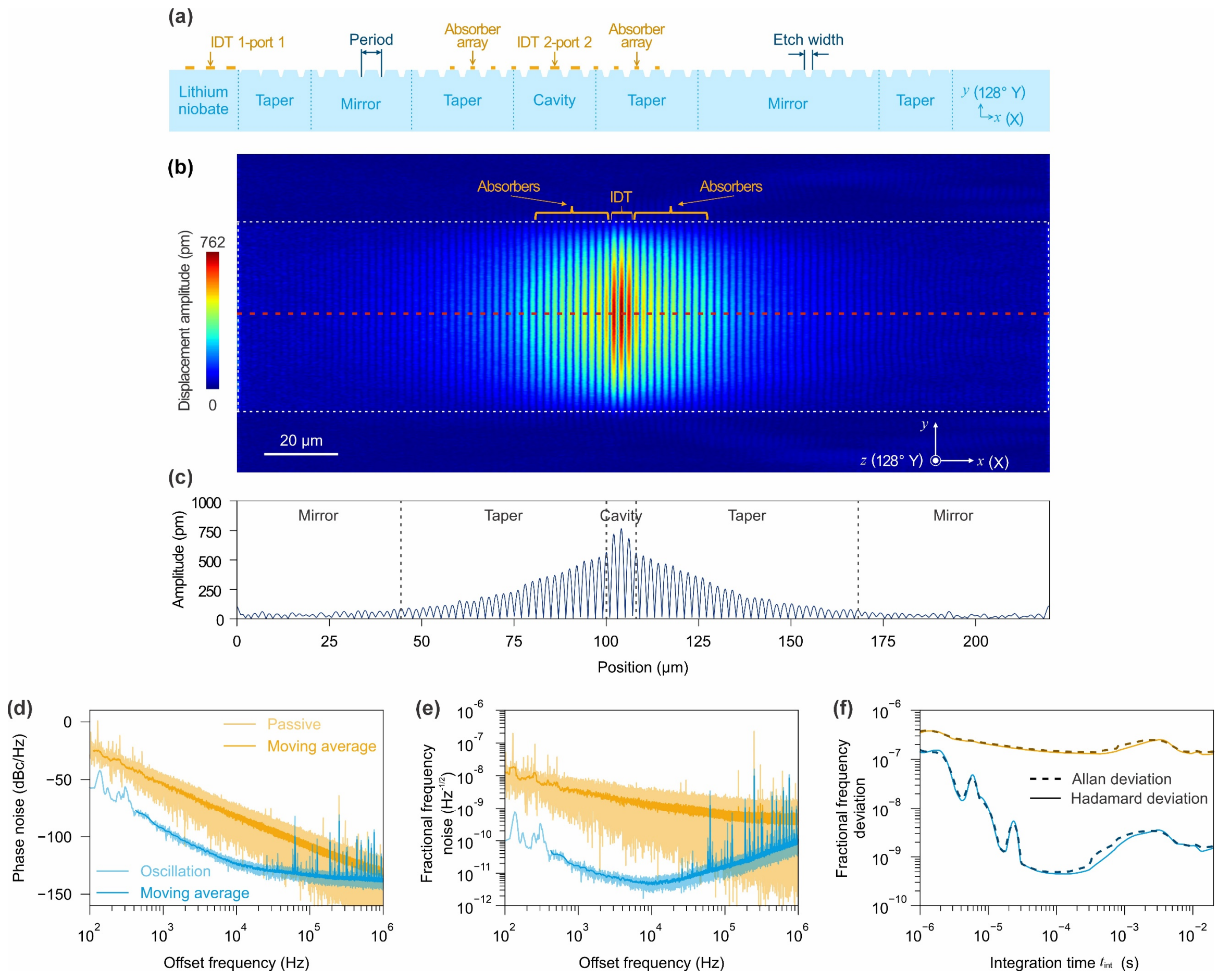

**FIG. 3. Design and characterization of the phononic crystal (PnC) resonant mode for Mid-infrared (MIR) detection.** (a) Cross-section schematic of PnC resonator. The PnC resonator is formed by a cavity segment sandwiched by two mirror segments with taper segments linking them. Grooves in cavity (mirrors) segments are with a spacing period of 2 μm (1.96μm). The spacing period of etched grooves is quadratically tapering from the cavity segment to the mirror segment. Mirrors transit to the unetched flat segments by linearly tapering the etch width. (b) Optically measured displacement field profile of the PnC mode. The color indicates displacement perpendicular to the device

surface. The PnC mode is excited by applying a continuous wave at the PnC-resonator-mode frequency applied at interdigital transducers 2 (IDT 2). (c) The cross-section of the displacement profile, indicated by the red dash line in (b). (d) Phase noise, (e) fractional frequency noise, (f) frequency deviation of our PnC resonator in the oscillation and passive measurement configurations. We use the definitions of Allan and Hadamard deviations such that white frequency noise has same deviation values.

We employ an oscillation configuration (**Fig. S3(b)**) to enable a direct and large-bandwidth readout of the PnC resonator frequency. The oscillation loop consists of our PnC resonator, a low noise amplifier, a phase shifter, and a coupler. The output of our oscillator is first downconverted by an in-phase / quadrature (I/Q) demodulator and then captured by an oscilloscope. Ensuring the accessibility for readers from different fields, we provide phase noise spectrum (**Fig. 3(d)**), fractional frequency noise spectrum (**Fig. 3(f)**), and Hadamard deviation (**Fig. 3(g)**) based on the same measured I/Q data. Our MIR detector oscillating at 1 GHz achieves a minimum fractional frequency noise of $4.32 \times 10^{-12}$ $\mathrm{Hz}^{-1/2}$ at offset frequency of 8,500 Hz, a minimum $\sigma\sqrt{t_{int}} = 3.7 \times 10^{-12}$ given by the frequency deviation $\sigma$ of $5.24 \times 10^{-10}$ at integration time $t_{int}$ of $5.0 \times 10^{-5}$ s. Compared to passive measurement (**Fig. S3(b)**), the oscillation configuration reduces the frequency noises by over two orders of magnitude (**Figs. 3(e) and 3(f)**). We note that similar observations showing the advantage of the oscillation configuration have been reported[40].

We attribute the ultralow frequency noise of our device for a few reasons. First, the surface acoustic wave (SAW)-based mechanical resonator is constructed on a bulk substrate, which has a much greater thermal conductivity than tethered (suspended) thin-film micromechanical resonators, leading to lower temperature fluctuation noise. Secondly, our design of PnC resonator defined by etched grooves[36-37] enables a higher *fQ* product than a typical suspended mechanical resonator or metal-defined SAW resonators. Additionally, the strong piezoelectricity of LN allows efficient coupling between mechanical and electrical fields, leading to a lower insertion of the oscillation loop and thus a lower noise. Focused discussion on the noise performance of the PnC oscillator is available in our previous work[37].

## 4. CHARACTERIZATION OF MIR DETECTION

We characterize responsivity, bandwidth, and effective areas of our oscillator-based MIR detector. Our experimental setup is shown in **Fig. S4**. The results of our devices with 6-and-9-μm dual-wavelength absorbers are shown in **Fig. 4**, while the results of our device with 6-μm single-wavelength absorbers are shown in **Fig. S5**. We use a quantum cascade laser at 6.30 μm and a distributed feedback laser at 9.55 μm as MIR inputs. The MIR inputs are modulated by a mechanical chopper with variable chopping frequency. The incident MIR power is calibrated to the power within the device area with absorbers (**Supplemental Note 3**).

When a 190-Hz-chopped 5.3-μW-power MIR beam at 6.30 μm wavelength is applied on our detector, we observe a frequency modulation of our oscillator-based detector (**Fig. 4(a)**). As lithium niobate has a negative temperature coefficient in elasticity, the oscillating frequency will reduce when light is on the detector. Our detector shows a time constant of 169 (164) μs in the frequency rise (fall) in response to the MIR light off (on) (**Fig. 4(b)**). By Fourier transform, we see the oscillating frequency of our detectors modulated at the chopper frequency and its higher order harmonics (**Fig. 4(c)**). Within this work, we define the response of our detector by the frequency modulation amplitude at the fundamental chopper frequency. For example, a frequency modulation 562 Hz induced by the MIR input of 5.3 μW (**Fig. 4(c)**).

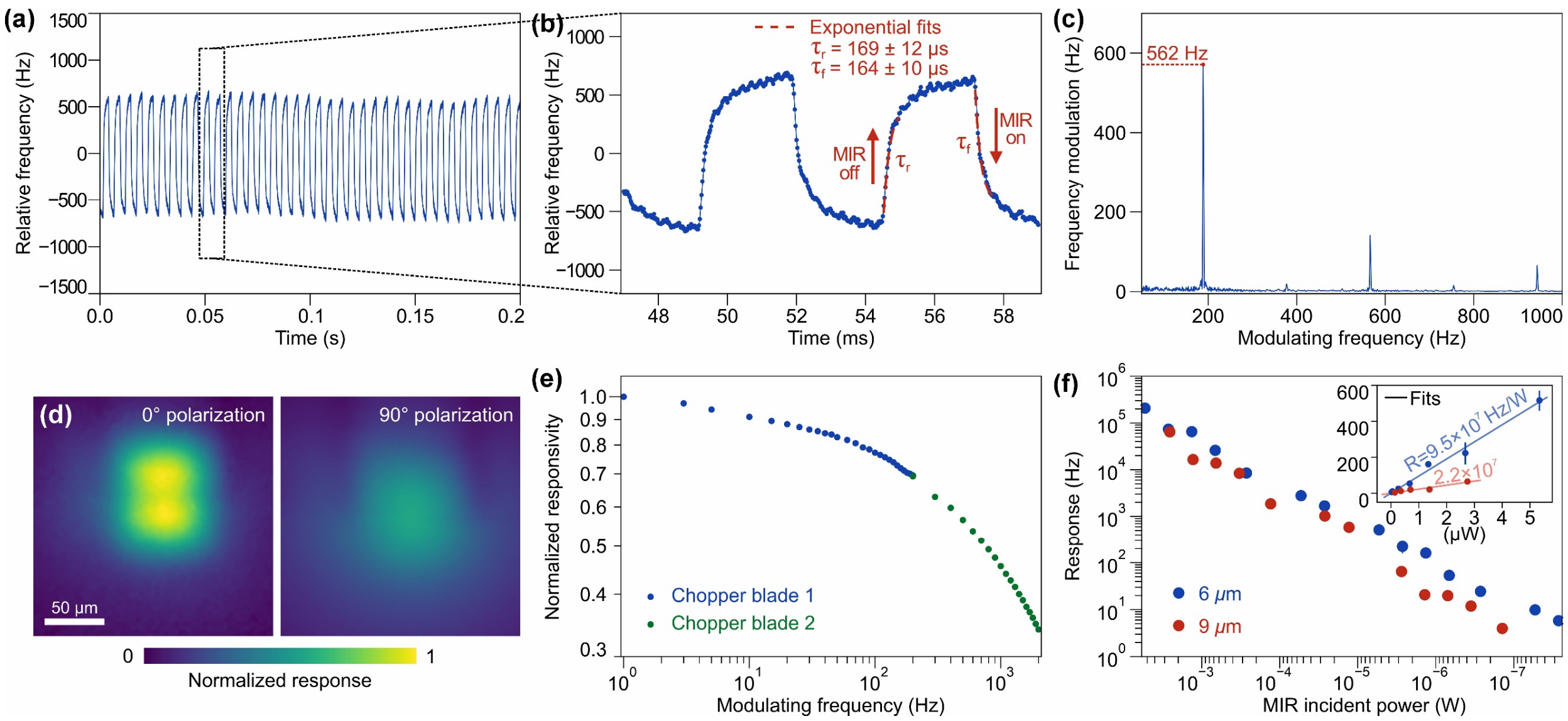

**FIG. 4. Characterization of our mid-infrared (MIR) detector with dual-wavelength absorbers.** (a) Measured modulation of oscillator frequency in response to a chopped MIR input. The incident MIR on the detector is with an average power of 5.3 μW chopped at 190 Hz. (b) Zoom in view of (a) showing the detector response time constants of 164 μs in when MIR input is turning on and 169 μs when turning off. (c) Frequency spectrum of the frequency modulation signal, showing a peak of 562 Hz at 190 Hz chopper frequency. (d) Response mapping of our detector to 0° and 90° polarized MIR input at 6.3μm. The mapping result is the convolution of the incident MIR beam with the effective area of our detector. (e) Frequency response of our detector at MIR incident power of 53.0 μW. Two chopper blades with different opening densities are used to cover the frequency range. 6-μm MIR light is used in (a)-(e). (f) Power dependent response of our detector at 6 and 9 μm wavelength. The incident MIR is modulated at 190 Hz. Inset: power dependent response in linear scale, and the linear fitting show the responsivities of our detector.

We map the response at different positions of our detector (**Fig. 4(d)**). We note that the mapping result is the convolution of the incident MIR beam with the effective area of our detector. Our detector shows a clear polarization dependent response. For incident light with polarization angle at 0° with respect to the antenna axis, the optical absorption of the antenna metasurface and LN beneath the antenna is enhanced near antenna resonance wavelengths; while for incident light with polarization angle at 90°, the optical absorption is mostly due to LN material absorption in the mid-IR spectral region. The measurement results confirmed an up to 250 % enhancement of infrared detection with antenna absorbers at 6.30 μm, which is around the 6 μm antenna resonance wavelength. As one of the IDTs is placed at the center of the acoustic resonator for optimized electromechanical coupling, the response is slightly lower at the center of the detector. This issue could be addressed by future designs that combine IDT electrodes and absorbers.

We characterize the frequency response of our detector by varying the chopper frequency (**Fig. 4(e)**). We measure a relative response of 0.77 at 100 Hz, 0.7 at 190 Hz, and 0.46 at 1 kHz, showing a frequency roll-off rate of 4.5 dB/decade. The frequency response is limited by the thermal dynamics of our detector. We note that the frequency roll-off of our detectors is much slower than the 20 dB/decade of RC-limited detectors. Therefore, the relationship between the fitted time constant and bandwidth is different from RC-limited detectors as well.

We test our detector over a large input power range up to 5.3 mW at 6.3 μm wavelength and 2.6 mW at 9 μm wavelength (**Fig. 4(f)**). Our detector features a large dynamic range of $10^7$, from its NEP of 197 pW/$\sqrt{\text{Hz}}$ (see discussion in **Sec. 5**) to 5.3 mW. We note that the maximum tested input power is limited by our MIR laser source and does not infer a device damage threshold. By fitting the measured responses at different incident MIR powers (inset of **Fig. 4(f)**), we extract a responsivity $R_{190\mathrm{Hz}}$ = 9.5×10$^{7}$ Hz/W at 6.3 μm MIR wavelength and 2.2×10$^{7}$ Hz/W at 9 μm wavelength, both at input modulating (chopper) frequency of 190 Hz. At near-DC modulating frequency, the responsivity $R_{\mathrm{DC}}$= 1.36×10$^{8}$ Hz/W and 3.14 ×10$^{7}$ Hz/W for 6.3 and 9 μm, respectively.

The detector with 6-μm single-wavelength absorbers shows similar performance (**Fig. S5**). The responsivity $R_{190\mathrm{Hz}}$ = 9.6×10$^{7}$ Hz/W at input modulating frequency of 190 Hz, and responsivity $R_{\mathrm{DC}}$= 1.37×10$^{8}$ Hz/W at near DC. Compared to near-DC (1 Hz) response, the relative response is 0.79 at 100 Hz, and 0.48 at 1 kHz modulating frequency. These results infer the compatibility of our device architecture with versatile absorbers for different wavelengths.

To demonstrate an application of the large dynamic range, we perform a scanning imaging of Tetris pattern using our detector (**Fig. 5**). The Tetris pattern card is mounted on a translational stage with a 3.28-mW 6.3-μm MIR laser as a light source (**Fig. 5(a)**). Using glass as attenuator, the transmission of Tetris patterns varies about six orders of magnitudes from 4.1×10$^{-6}$ (bottom-right "O" Tetris piece) to 1 (bottom-left "S" Tetris piece) (**Fig. 5(b)**). The scanning results (**Figs. 5(c) and 5(d)**) clearly show the induced oscillation frequency shifts, ranging from 3 Hz to 266 kHz, at different patterns. The spatial resolution of our scanning image is limited by the spot size at the pattern card, which is about 2 millimeters in diameter.

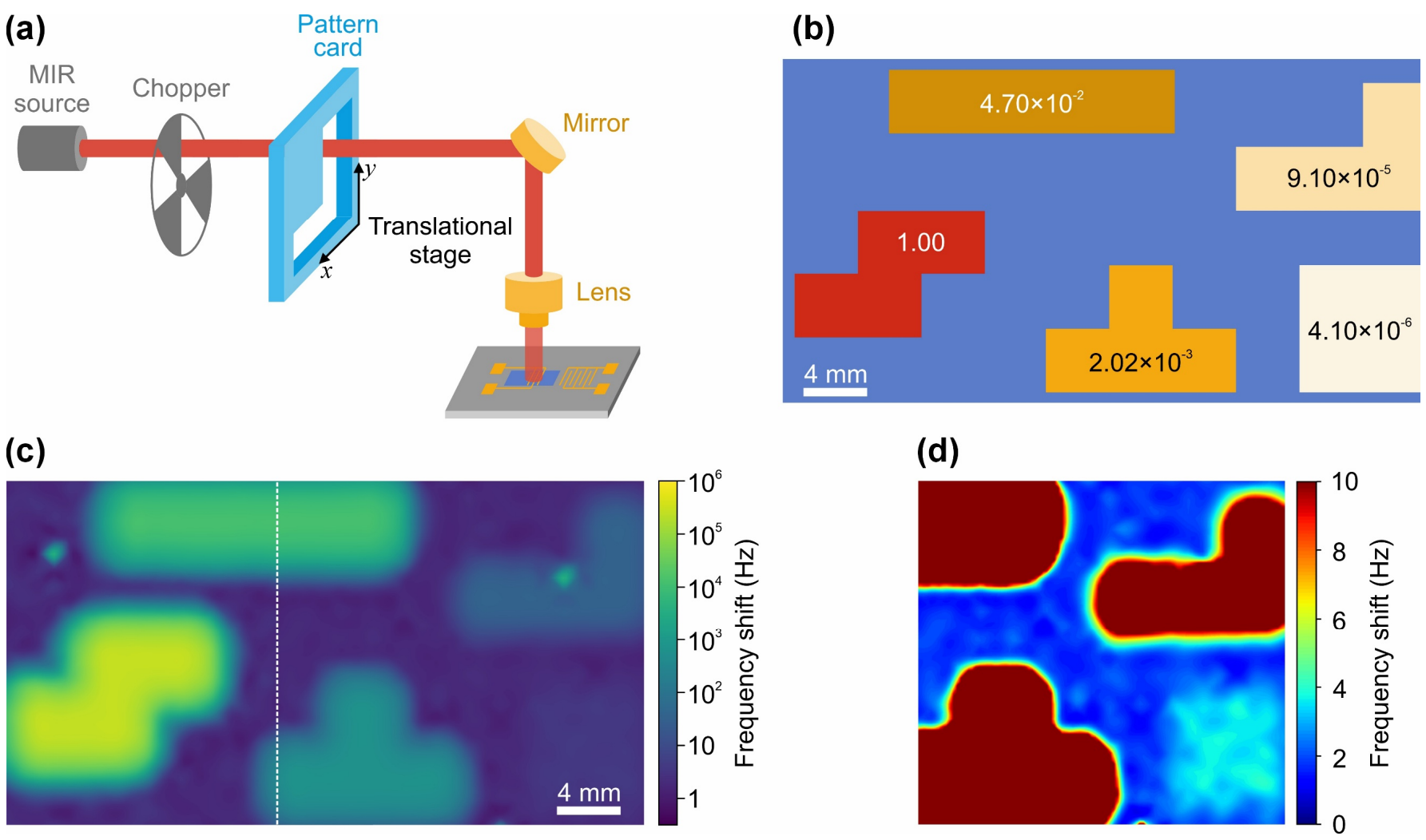

**FIG. 5. Scanning imaging of Tetris patterns with orders-of-magnitude difference in MIR power.** (a) Illustration of the scanning imaging setup. (b) Tetris patterns with different transmissions. The pattern is prepared by a 3D printed frame covered with glass as attenuators. Transmissions are marked at each opening. (c) Scanning image of the patterns. The colors show the induced frequency shift of our detector in the log scale. Dashed lines indicate the stitching of two scans due to the moving range of the translational stage. (d) Scanning image of the pattern plotted in the linear scale. The data in (c) and (d) are identical but plotted in different scales. The incident MIR is at 6.3 μm and 3.28 mW at detector without attenuation.

## 5. Noise Equivalent Power and Comparison

The frequency-dependent NEP of a detector can be given by $f_0\sqrt{S_y(f)}/R(f)$, where $S_y(f)$ in Hz$^{-1}$ is the power spectral density of fractional frequency noise, $f_0$ is oscillation frequency, and $R(f)$ in Hz/W is the responsivity at input frequency of $f$. By this definition, we extract the frequency-dependent NEP (Fig. 6) using the frequency noise in Fig. 3(e) and frequency-dependent response in Fig. 4(e). With respect to the incident light on the detector effective area, our dual-wavelength detector features a minimum NEP of 197 pW/$\sqrt{\text{Hz}}$ at input frequency of 1.8 kHz at 6.3 μm input wavelength. At low input modulating frequencies (below 500 Hz), the NEP is limited by the larger frequency noise in long time scale. At the input frequency within the range of 700 Hz to 2 kHz, the NEP is almost flat, since both the responsivity and the frequency noise reduce with the increasing frequency. On the other hand, if we define the detection bandwidth by the frequency-dependent NEP, our detector shows a bandwidth of at least 2 kHz.

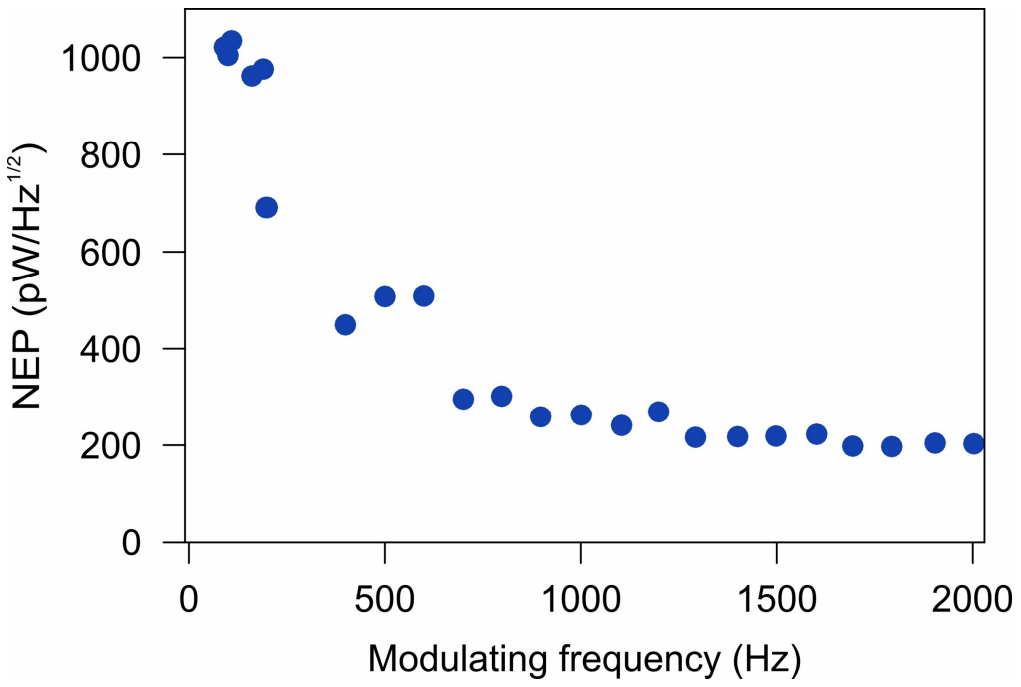

**FIG. 6. Frequency-dependent noise equivalent power (NEP) of our mid-infrared detector at 6.3 μm wavelength.**

**Figure 7** summarizes the performance of MIR detectors at room temperature; incident powers are calculated using MIR power on the detector effective area. While MIR photodiodes have demonstrated large bandwidths of 7 GHz based on semiconductor[11] and 500 GHz based on two-dimensional materials[20], the mechanical resonator-based approaches show advantages in lower NEP at longer MIR wavelength (**Fig. 7(a)**). Among mechanical-resonator-based detectors, our fully supported detectors show competitive NEP and time constants, leading to a competitive figure-of-merit, NEP· τ (**Fig. 7(b)** and **Table S1**). We note that due to the availability of data in previous works and products, are used in this comparison.

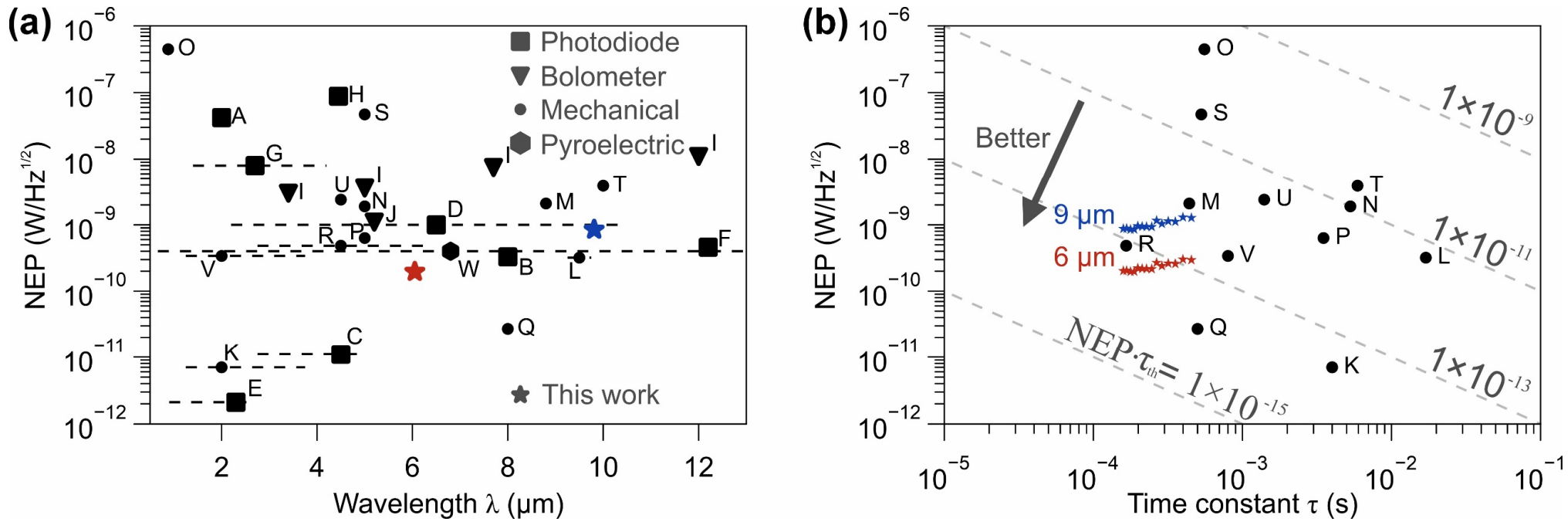

**FIG. 7. Figure of merits for MIR detectors.** (a) NEP vs detection wavelengths among photodiodes, bolometers, and mechanical resonators. The plots show the experimentally demonstrated results. If a broadband thermal source is used, the peak detection wavelength is determined by the source temperature or the fiber/filters used. Dashed lines indicate the operation wavelength range, if data is available. (b) NEP vs detector time constant among mechanical based MIR detectors. Labels in the figure: A[11], B: Thorlabs, VML8T0 [41], C: Thorlabs, VL5T0 [42], D: Thorlabs, VML10T0 [43], E: Thorlabs, PDA10DT [44], F[45], G[20], H[16], I[28], J[29], K[31], L[46], M[47], N[48], O[49], P[50], Q[33], R[51], S[52], T[53], U[54], V[55], W: Laser Components, RP NO1[56]. When frequency dependent NEP data is not availability for previous works or products, a noise-floor-limited NEP $= f_0\sigma\sqrt{t_{int}}/R_{\mathrm{DC}}$ was used, which could overestimate the performance.

We model the dynamics of our oscillator-based MIR detector by its thermodynamics and mechanical oscillation (**Supplemental Note 4**). The temperature change of the resonator induced by the MIR absorption is parametrically coupled to the resonant frequency of the mechanical resonator. The thermodynamics limits the detector response time. Thermodynamic bandwidth could be further improved by employing thin-film lithium niobate on sapphire[57]. We consider two main noise sources of our oscillator-based MIR detector: temperature fluctuation noise of the mechanical resonator and the thermal noise in the oscillation loop. The temperature fluctuation noise is considered a fundamental limit for mechanical-resonator-based MIR detection, which imposes a NEP of 4.38 $\mathrm{pW}/\sqrt{\mathrm{Hz}}$ for our devices, which could be improved by employing a smaller resonator. The thermal noise in the oscillation loop that amplified by the low-noise amplifier imposes a NEP of 0.62 $\mathrm{pW}/\sqrt{\mathrm{Hz}}$. Thus, the temperature fluctuation noise is the dominating noise of our detectors. We note that the NEP from the theoretical model is lower than our experimental demonstration, and we speculate a few possible reasons for the mismatch. (1) the low noise amplifier is working in the saturation region for our oscillation detector, leading to a higher noise factor, and (2) the rate equations in our toy model may be oversimplified to describe the nonlinear dynamics and noise performance. Despite such mismatch, our analytical solutions and simulations can still provide insights into future developments. Looking into the future, a higher $Q$ factor, a lower insertion loss, or a higher power level in the oscillation loop could further improve the NEP.

## 6. CONCLUSION AND OUTLOOK

Our mechanical-oscillator-based MIR detectors feature ultra-low noise, high sensitivity, fast response, and large dynamic ranges without any cooling at room temperature. Compared to passive readout approaches, the oscillation configuration of our detector enables an ultralow frequency noise in reading out the resonant frequency. Integrating plasmonic antennas metasurface absorbers with these microscale MIR detectors/detector arrays could provide versatile, compact and high speed solutions for single or multi-wavelength, spectroscopic and polarization detection over the whole infrared wavelength range, which find various applications in vibrational spectroscopy[58], infrared polarimetry[59-60], remote sensing[61-63], chemical analysis[64-65], and biomedical diagnosis[66]. With flexible absorber designs, our MIR detector architecture can also enable monolithic solutions for UV-FIR applications. Moreover, its fabrication process is compatible with recently established lithium niobate foundries and scalable at wafer-scale[67]. For integrated platforms, the PnC resonators can be integrated with optical waveguides[68] enabling sensitive detection for MIR photonic integrate circuits.

## FUNDING

DOE Office of Science User Facility (CNMS2022-B-01473, CNMS2024-B-02643); Defense Advanced Research Projects Agency (DARPA) (HR00112320031).

## ACKNOWLEDGMENTS

Device fabrication was conducted as part of user projects at the Center for Nanophase Materials Sciences (CNMS), which is a DOE Office of Science User Facility. This work is supported by the Defense Advanced Research Projects Agency (DARPA) OPTIM program. The views and conclusions contained in this document are those of the authors and do not necessarily reflect the position or the policy of the Government. No official endorsement should be inferred. Approved for public release; distribution is unlimited.

## DISCLOSURES

Virginia Tech and Arizona State University filed a provisional application for patent related to this work.
The authors declare no other competing interests.

## DATA AVAILABILITY STATEMENT

Data that supports the findings of this study are available from the corresponding authors upon reasonable request.

# Reference

[1] ISO 20473:2007, Optics and photonics -- Spectral bands

[2] S. F. Hönig, M. Kishimoto, P. Gandhi, A. Smette, D. Asmus, W. Duschl, M. Polletta, G. Weigelt, *Astron. Astrophys.* **2010**, 515, A23.

[3] M. Yu, Y. Okawachi, A. G. Griffith, N. Picqué, M. Lipson, A. L. Gaeta, *Nat. Commun.* **2018**, 9, 1869.

[4] M. Liu, R. M. Gray, L. Costa, C. R. Markus, A. Roy, A. Marandi, *Nat. Commun.* **2023**, 14, 1044.

[5] E. F. J. Ring, K. Ammer, *Physiological Measurement* **2012**, 33, R33.

[6] D. Rodrigo, O. Limaj, D. Janner, D. Etezadi, F. J. García de Abajo, V. Pruneri, H. Altug, *Science* **2015**, 349, 165.

[7] R. Ishimwe, K. Abutaleb, F. Ahmed, *Advances in remote Sensing* **2014**, 3, 128.

[8] X. Pang, O. Ozolins, S. Jia, L. Zhang, R. Schatz, A. Udalcovs, V. Bobrovs, H. Hu, T. Morioka, Y.-T. Sun, J. Chen, S. Lourdudoss, L. K. Oxenløwe, S. Popov, X. Yu, *J. Lightwave Technol.* **2022**, 40, 3149.

[9] P. D. Anderson, J. D. Beck, W. Sullivan, C. Schaake, J. McCurdy, M. Skokan, P. Mitra, X. Sun, *J. Electron. Mater.* **2022**, 51, 6803.

[10] J. Beck, C. Wan, M. Kinch, J. Robinson, P. Mitra, R. Scritchfield, F. Ma, J. Campbell, *J. Electron. Mater.* **2006**, 35, 1166.

[11] D. Chen, S. D. March, A. H. Jones, Y. Shen, A. A. Dadey, K. Sun, J. A. McArthur, A. M. Skipper, X. Xue, B. Guo, J. Bai, S. R. Bank, J. C. Campbell, *Nat. Photonics* **2023**, 17, 594.

[12] A. H. Jones, S. D. March, S. R. Bank, J. C. Campbell, *Nat. Photonics* **2020**, 14, 559.

[13] C. Xie, M. Aziz, V. Pusino, A. Khalid, M. Steer, I. G. Thayne, M. Sorel, D. R. S. Cumming, *Optica* **2017**, 4, 1498.

[14] A. Delga, in *Mid-infrared Optoelectronics*, (Eds: E. Tournié, L. Cerutti), Woodhead Publishing, 2020.

[15] B. F. Levine, *J. Appl. Phys.* **1993**, 74, R1.

[16] Y. Yao, R. Shankar, P. Rauter, Y. Song, J. Kong, M. Loncar, F. Capasso, *Nano Lett.* **2014**, 14, 3749.

[17] Q. Guo, A. Pospischil, M. Bhuiyan, H. Jiang, H. Tian, D. Farmer, B. Deng, C. Li, S.-J. Han, H. Wang, Q. Xia, T.-P. Ma, T. Mueller, F. Xia, *Nano Lett.* **2016**, 16, 4648.

[18] X. Chen, X. Lu, B. Deng, O. Sinai, Y. Shao, C. Li, S. Yuan, V. Tran, K. Watanabe, T. Taniguchi, D. Naveh, L. Yang, F. Xia, *Nat. Commun.* **2017**, 8, 1672.

[19] Y. Fang, Y. Ge, C. Wang, H. Zhang, *Laser Photonics Rev.* **2020**, 14, 1900098.

[20] S. M. Koepfli, M. Baumann, Y. Koyaz, R. Gadola, A. Güngör, K. Keller, Y. Horst, S. Nashashibi, R. Schwanninger, M. Doderer, E. Passerini, Y. Fedoryshyn, J. Leuthold, *Science* **2023**, 380, 1169.

[21] S. M. M. Friis, L. Høgstedt, *Opt. Lett.* **2019**, 44, 4231.

[22] T. W. Neely, L. Nugent-Glandorf, F. Adler, S. A. Diddams, *Opt. Lett.* **2012**, 37, 4332.

[23] P. J. Rodrigo, L. Høgstedt, S. M. M. Friis, L. R. Lindvold, P. Tidemand-Lichtenberg, C. Pedersen, *Laser Photonics Rev.* **2021**, 15, 2000443.

[24] S. Wolf, J. Kiessling, M. Kunz, G. Popko, K. Buse, F. Kühnemann, *Opt. Express* **2017**, 25, 14504.

[25] P. Tidemand-Lichtenberg, J. S. Dam, H. V. Andersen, L. Høgstedt, C. Pedersen, *J. Opt. Soc. Am. B* **2016**, 33, D28.

[26] Y. Cai, Y. Chen, X. Xin, K. Huang, E. Wu, *Photon. Res.* **2022**, 10.

[27] J. Fang, K. Huang, R. Qin, Y. Liang, E. Wu, M. Yan, H. Zeng, *Nat. Commun.* **2024**, 15, 1811.

[28] S. Yuan, R. Yu, C. Ma, B. Deng, Q. Guo, X. Chen, C. Li, C. Chen, K. Watanabe, T. Taniguchi, F. J. García de Abajo, F. Xia, *ACS Photonics* **2020**, 7, 1206.

[29] J. Goldstein, H. Lin, S. Deckoff-Jones, M. Hempel, A.-Y. Lu, K. A. Richardson, T. Palacios, J. Kong, J. Hu, D. Englund, *Nat. Commun.* **2022**, 13, 3915.

[30] J. Shim, J. Lim, I. Kim, J. Jeong, B. H. Kim, S. K. Kim, D.-M. Geum, S. Kim, *arXiv e-prints* **2024**, 2405.14155.

[31] M. Piller, J. Hiesberger, E. Wistrela, P. Martini, N. Luhmann, S. Schmid, *IEEE Sens. J.* **2023**, 23, 1066.

[32] M. E. Gülseren, M. Benson, R. W. Parker, J. Segovia-Fernandez, E. T. T. Yen, J. S. Gómez-Díaz, *IEEE Sens. J.* **2024**, 24, 17313.

[33] L. Laurent, J.-J. Yon, J.-S. Moulet, M. Roukes, L. Duraffourg, *Physical Review Applied* **2018**, 9, 024016.

[34] S.-E. Stanca, V. R. Rayapati, A. Chakraborty, J. Dellith, W. Fritzsche, G. Zieger, H. Schmidt, *Sci. Rep.* **2024**, 14, 22709.

[35] J. A. Bossard, L. Lin, S. Yun, L. Liu, D. H. Werner, T. S. Mayer, *ACS Nano* **2014**, 8, 1517.
[36] L. Shao, S. Maity, L. Zheng, L. Wu, A. Shams-Ansari, Y.-I. Sohn, E. Puma, M. N. Gadalla, M. Zhang, C. Wang, E. Hu, K. Lai, M. Lončar, *Physical Review Applied* **2019**, 12, 014022.
[37] Z. Xi, J. G. Thomas, J. Ji, D. Wang, Z. Cen, I. I. Kravchenko, B. R. Srijanto, Y. Yao, Y. Zhu, L. Shao, *Physical Review Applied* **2025**, 23, 024054.
[38] R. B. Ward, *IEEE Transactions on Ultrasonics, Ferroelectrics, and Frequency Control* **1990**, 37, 481.
[39] J. Thomas, Z. Xi, J. Ji, Jun, G. Shi, B. Srijanto, I. Kravchenko, Ivan, et al. (2025). High Speed Surface Acoustic Wave Imaging with Spectral Interferometry. Optica Open. Preprint. https://doi.org/10.1364/opticaopen.28477103.v1
[40] Y. Huang, J. G. Flor Flores, Y. Li, W. Wang, D. Wang, N. Goldberg, J. Zheng, M. Yu, M. Lu, M. Kutzer, D. Rogers, D.-L. Kwong, L. Churchill, C. W. Wong, *Laser Photonics Rev.* **2020**, 14, 1800329.
[41] https://www.thorlabs.com/thorproduct.cfm?partnumber=VML8T0
[42] https://www.thorlabs.com/thorproduct.cfm?partnumber=VL5T0
[43] https://www.thorlabs.com/thorproduct.cfm?partnumber=VML10T0
[44] https://www.thorlabs.com/thorproduct.cfm?partnumber=PDA10DT
[45] Q. Guo, R. Yu, C. Li, S. Yuan, B. Deng, F. J. García de Abajo, F. Xia, *Nat. Mater.* **2018**, 17, 986.
[46] P. Markus, L. Niklas, C. Miao-Hsuan, S. Silvan, presented at *Proc.SPIE*, **2019**.
[47] Y. Hui, J. S. Gomez-Diaz, Z. Qian, A. Alù, M. Rinaldi, *Nat. Commun.* **2016**, 7, 11249.
[48] Y. Hui, S. Kang, Z. Qian, M. Rinaldi, *Journal of Microelectromechanical Systems* **2021**, 30, 165.
[49] V. J. Gokhale, M. Rais-Zadeh, *Journal of Microelectromechanical Systems* **2014**, 23, 803.
[50] Z. Qian, S. Kang, V. Rajaram, M. Rinaldi, presented at *2016 IEEE SENSORS*, 30 Oct.-3 Nov. 2016, **2016**.
[51] Z. Qian, V. Rajaram, S. Kang, M. Rinaldi, *Appl. Phys. Lett.* **2019**, 115, 261102.
[52] Z. Qian, Y. Hui, F. Liu, S. Kang, S. Kar, M. Rinaldi, *Microsystems & Nanoengineering* **2016**, 2, 16026.
[53] M. B. Pisani, K. Ren, P. Kao, S. Tadigadapa, *Journal of Microelectromechanical Systems* **2011**, 20, 288.
[54] Y. Hui, M. Rinaldi, presented at *2013 Transducers & Eurosensors XXVII: The 17th International Conference on Solid-State Sensors, Actuators and Microsystems (TRANSDUCERS & EUROSENSORS XXVII)*, 16-20 June 2013, **2013**.
[55] W. C. Ang, P. Kropelnicki, H. Campanella, Y. Zhu, A. B. Randles, H. Cai, Y. A. Gu, K. C. Leong, C. S. Tan, presented at *2014 IEEE 27th International Conference on Micro Electro Mechanical Systems (MEMS)*, 26-30 Jan. 2014, **2014**.
[56] https://www.lasercomponents.com/us/product/pr-series-ir-thz-receivers/
[57] L. Shao, S. W. Ding, Y. Ma, Y. Zhang, N. Sinclair, M. Lončar, *Physical Review Applied* **2022**, 18, 054078.
[58] B. Zheng, H. Zhao, B. Cerjan, S. Yazdi, E. Ringe, P. Nordlander, N. J. Halas, *Applied Physics Letters* **2018**, 113.
[59] M. Dai, C. Wang, B. Qiang, F. Wang, M. Ye, S. Han, Y. Luo, Q. J. Wang, *Nature Communications* **2022**, 13, 4560.
[60] J. Chen, F. Yu, X. Liu, Y. Bao, R. Chen, Z. Zhao, J. Wang, X. Wang, W. Liu, Y. Shi, C.-W. Qiu, X. Chen, W. Lu, G. Li, *Light: Science & Applications* **2023**, 12, 105.
[61] S. Yang, X. Yan, H. Qin, Q. Zeng, Y. Liang, H. Arguello, X. Yuan, *Remote Sensing* **2021**, 13, 741.
[62] P. Y. Deschamps, F. M. Breon, M. Leroy, A. Podaire, A. Bricaud, J. C. Buriez, G. Seze, *IEEE Transactions on Geoscience and Remote Sensing* **1994**, 32, 598.
[63] X. Tan, H. Zhang, J. Li, H. Wan, Q. Guo, H. Zhu, H. Liu, F. Yi, *Nature Communications* **2020**, 11, 5245.
[64] D. Knez, B. W. Toulson, A. Chen, M. H. Ettenberg, H. Nguyen, E. O. Potma, D. A. Fishman, *Science Advances* **2022**, 8, eade4247.
[65] L. Shi, X. Liu, L. Shi, H. T. Stinson, J. Rowlette, L. J. Kahl, C. R. Evans, C. Zheng, L. E. P. Dietrich, W. Min, *Nature Methods* **2020**, 17, 844.
[66] Y. Zhao, S. Kusama, Y. Furutani, W.-H. Huang, C.-W. Luo, T. Fuji, *Nature Communications* **2023**, 14, 3929.
[67] K. Luke, P. Kharel, C. Reimer, L. He, M. Loncar, M. Zhang, *Opt. Express* **2020**, 28, 24452.
[68] L. Shao, S. W. Ding, N. Sinclair, J. G. Leatham, M. Loncar, presented at *Conference on Lasers and Electro-Optics*, San Jose, California, 2022/05/15, **2022**.

## Supplemental Figures

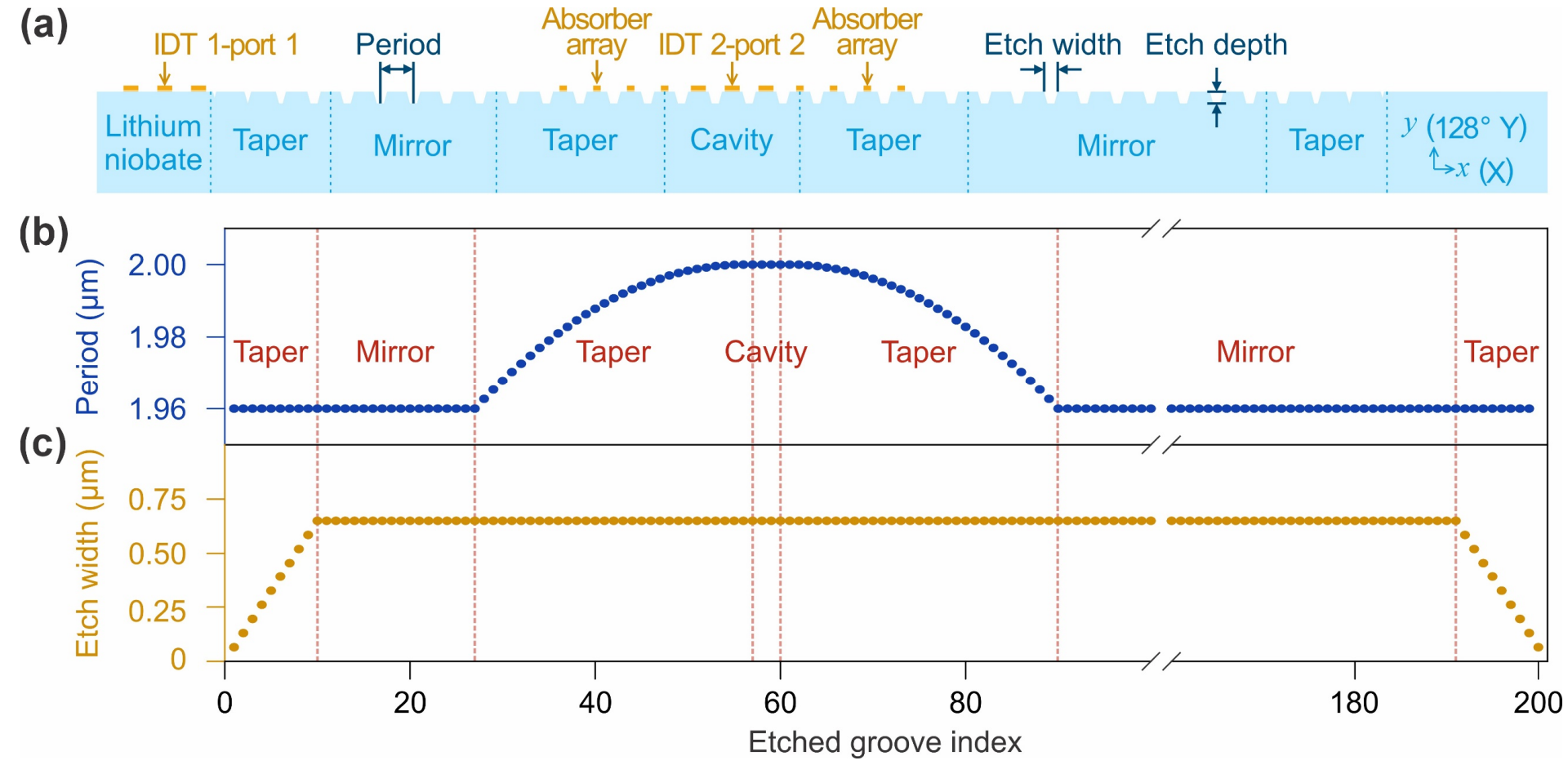

**Fig. S1 | Design of our phononic crystal (PnC) resonator.** (a) Cross-section schematic of our PnC resonator. The PnC resonator is defined by a series of etched grooves with different periods in different segments. (b) Pitch and (c) etch width of each etched grooves. The pitch between two adjacent etched grooves in the cavity region is 2.00 µm and quadratically tapers to 1.96 µm in the mirror regions. The etch widths of the first 10 grooves near interdigital transducer 1 (IDT 1) linearly taper to 0.65 µm in the mirror region. The etch depth of grooves is 100 nm. Counted along the $+x$ direction, the number of grooves in different regions is: 10 (taper), 16 (mirror), 30 (taper), 4 (cavity), 30 (taper), 100 (mirror), 10 (taper).

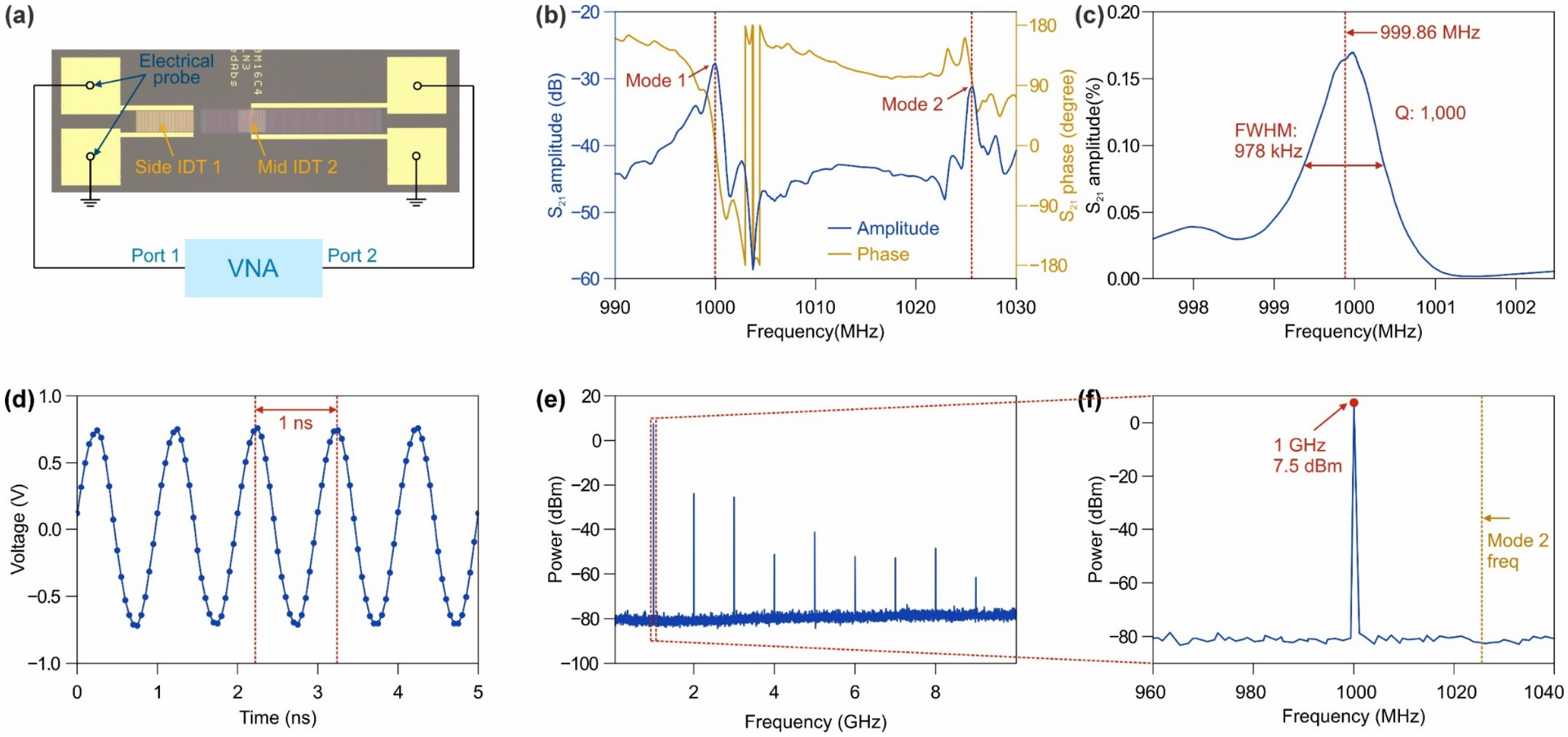

**Fig. S2 | Characterization of phononic crystal (PnC) resonator mode and surface acoustic wave (SAW) oscillator signal.** (a) Experimental setup. A vector network analyzer (VNA, Keysight P5004A) is used to measure the PnC resonator mode. The side (mid) interdigital transducers (IDT) are connected to Port 1 (Port 2) of VNA via electrical probes and 50-Ω coaxial cables. The losses due to cables are calibrated out in the measurements. (b) Measured S21 spectrum of the PnC resonator. The PnC resonator shows two resonance modes. The interested PnC mode locating at around 1 GHz shows a peak with transmission of -28 dB and is utilized to build the self-oscillation system. (c) Measured amplitude of transmission S21 spectra near the PnC mode. The PnC mode is centered at 999.86 MHz with full width half maximum (FWHM) of 978 kHz, resulting in a quality (Q) factor of 1,000. (d) Measured time-domain waveform of the oscillator signal, which is directly captured by a high-resolution oscilloscope (Rohde & Schwarz, RTO6) at sampling rate of 20 GSa/s. (e) Measured frequency spectrum of oscillator signal and (f) zoom-in view of frequency spectrum around 1 GHz.

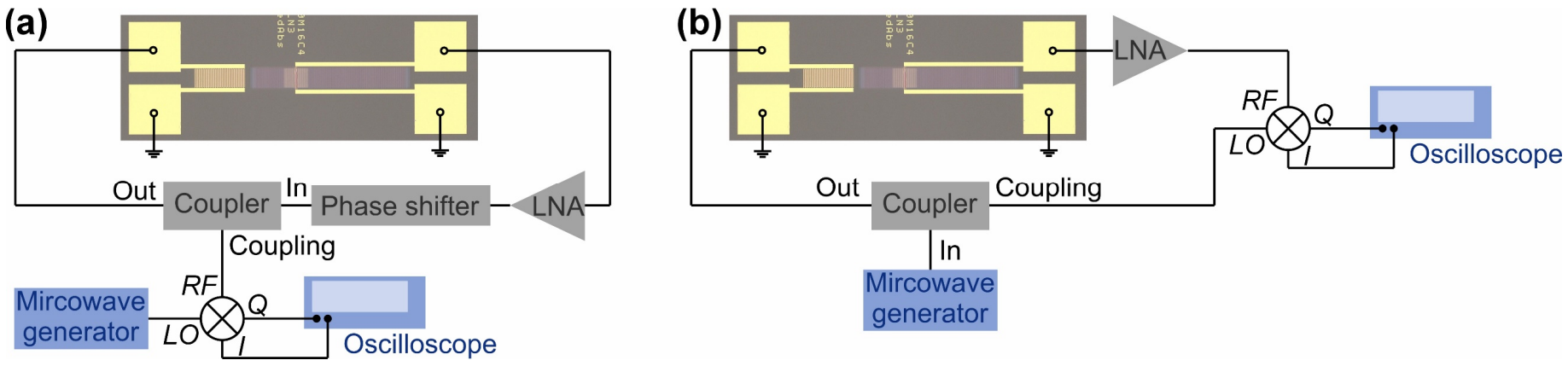

**Fig. S3 |** (a) Self-oscillation system based on the phononic crystal (PnC) resonator. The system consists of the PnC resonator, a low noise amplifier (LNA), a mechanical phase shifter, and a microwave coupler. An in-phase / quadrature (I/Q) demodulator, an ultra-low noise microwave generator and a high-resolution oscilloscope are used for readout of the oscillation signal coupled out from the coupler. Oscilloscope captures the I/Q data from the I/Q demodulator output ports for further data processing. (b) Experimental setup for the frequency deviation measurement of the passive resonator shown in Fig. 3(d).

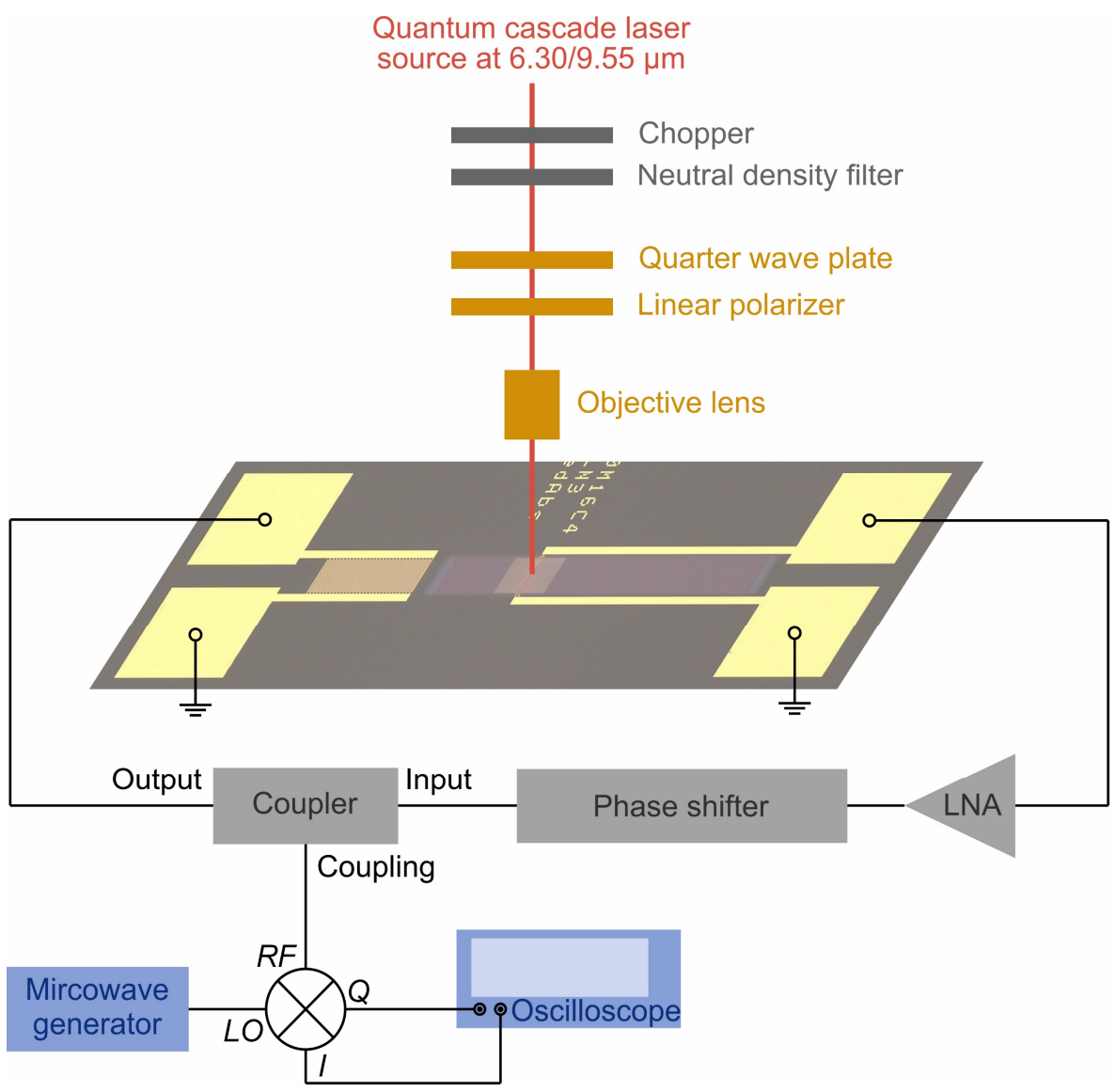

**Fig. S4 | Experimental setup for mid-infrared (MIR) detection.** A MIR tunable quantum cascade laser (working wavelength at 6.30 μm, Midcat-1100, Daylight solution) / MIR distributed feedback laser (working wavelength at 9.55 um, QD9550C2, Thorlabs) is used as source for 6/9-μm MIR detection measurements. The MIR laser chopped by an optical chopper is used as the input signal incident onto the detector. A series of neutral density (ND) filters (#12-003 to #12-017, Edmund optics and NDIR03B, Thorlabs) are used as attenuators to provide different incident power onto the detector. A low-order quarter waveplate (WPLQ05M-4500, Thorlabs) and a linear polarizer (WP25H-B, Thorlabs) are used to tune the polarization of the MIR laser. The chopped MIR is focused on the detector by a MIR objective lens (#3423, Edmund optics).

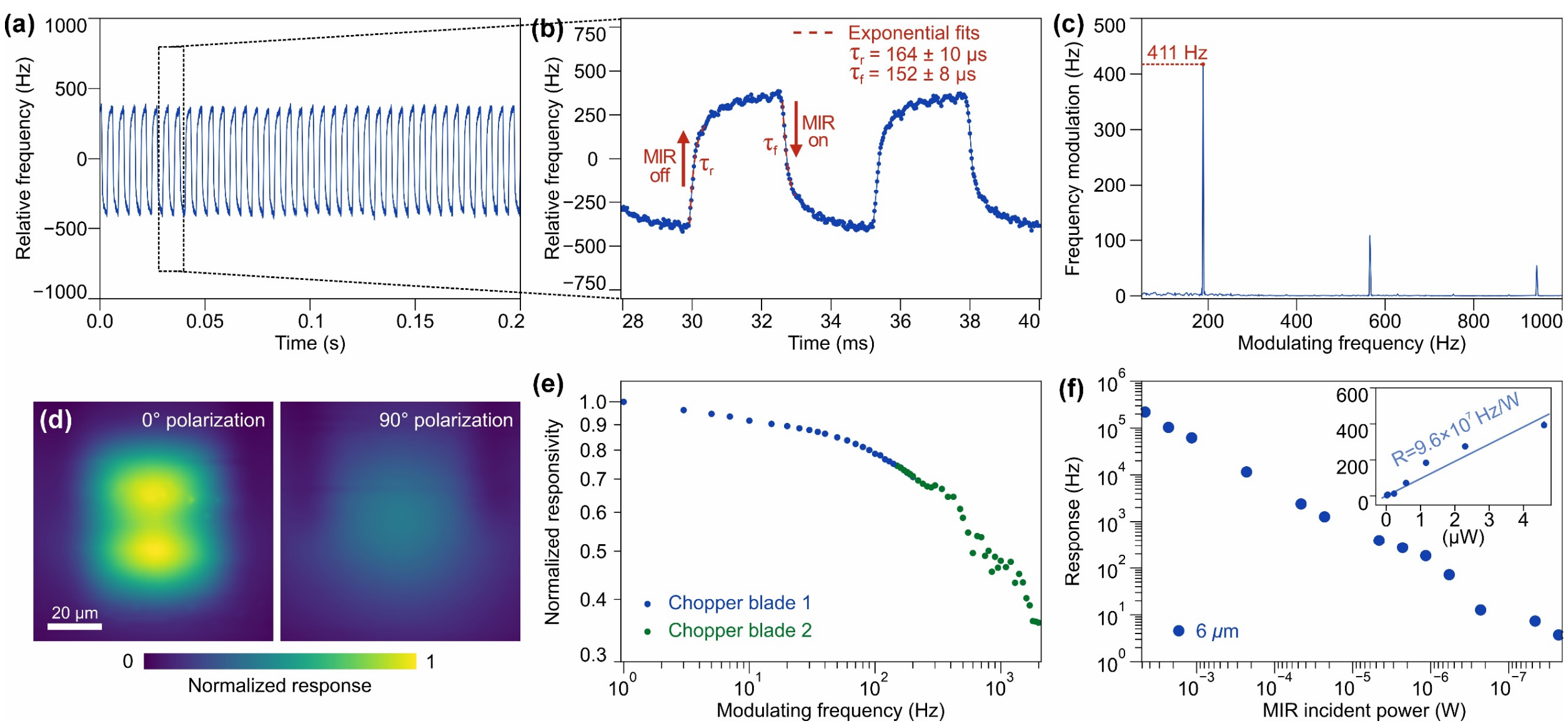

**Fig. S5 | Characterization of our 6-μm mid-infrared (MIR) detector.** (a) Measured modulation of oscillator frequency in response to a chopped MIR input. The incident MIR on the detector is with an average power of 4.55 μW chopped at 190 Hz. (b) Zoom in view of (a) showing the detector response time constant of 164 μs in when MIR input is turning on and 152 μs when turning off. (c) Frequency spectrum of the frequency modulation signal, showing a peak of 411 Hz at 190-Hz chopper frequency. (d) Response mapping of our detector to 0° and 90° polarized MIR input. The bright area is the result of the convolution between the temperature change induced by the MIR input spot and the phononic crystal mode. (e) Frequency response of our detector at MIR incident power of 45.5 μW. Two chopper blades with different opening densities are used to cover the frequency range. (f) Power dependent response of our detector. The incident MIR is modulated at 190 Hz. Inset: power dependent response in linear scale, and the linear fitting show the responsivity of our detector.

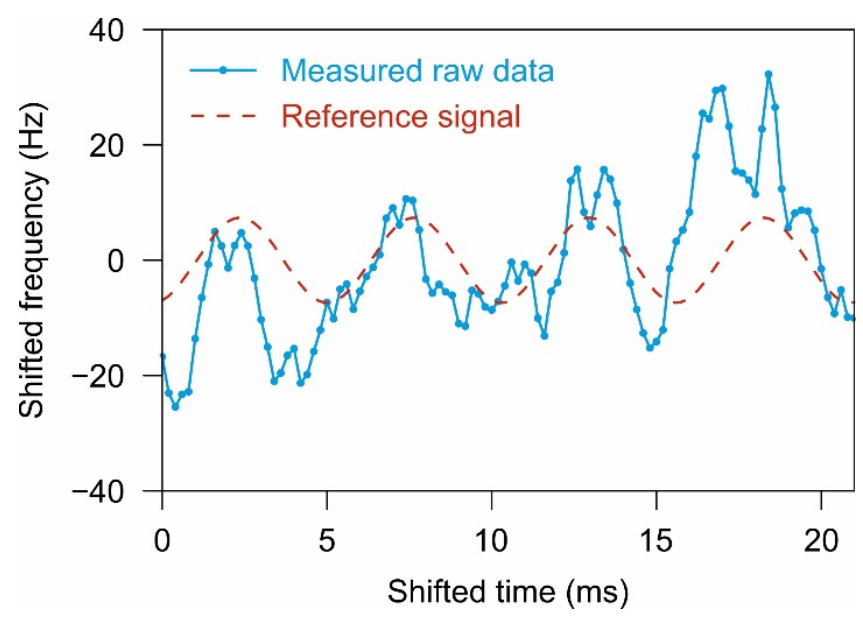

**Fig. S6 | Measured oscillator frequency in time domain.** The mid-infrared (MIR) detector is frequency modulated (the modulating frequency is 190 Hz) by a 6-μm laser with input power of 27 nW. Blue dotted line is measured oscillator frequency with sampling rate of 5,000 Hz (i.e. 20 ms sampling interval). Red dashed line is the reference signal with frequency of ~190 Hz. The amplitude and phase information of the reference signal is extracted from the Fourier frequency spectrum of the time-domain signal (total acquisition time is 1 second).

## Table S1 Comparison of MIR detectors based on mechanical resonator

| Label in Fig. 6(b) | Structure | Material of mechanical resonator | NEP [1] ($W/\sqrt{Hz}$) | Time constant $\tau$ (s) | NEP·$\tau$ ($s^{3/2} \cdot W$) | Detected peak wavelength (μm) | Demonstrated wavelength range (μm) | Mechanical resonant frequency (MHz) | Relative frequency noise ($Hz^{-1/2}$) | Responsivity ($W^{-1}$) | Effective area ($\mu m^2$) | D* ($cm \cdot \sqrt{Hz}/W$) |
|---|---|---|---|---|---|---|---|---|---|---|---|---|
| This work (6 μm) | **Bulk** | LN | 3.10E-11 | 1.69E-04 | **5.21E-15** | 6.3 | | 1,000 | **4.32E-12** | 0.137 | 48 × 56 | 1.67E+08 |
| This work (9 μm) | **Bulk** | LN | 1.55E-10 | 1.69E-04 | 2.62E-14 | 9.55 | | 1,000 | **4.32E-12** | 0.031 | 48 × 56 | 3.34E+07 |
| K | Suspended | SiN | **7.00E-12** | 4.00E-03 | 2.80E-14 | 2 | 1.25~3.75 | 0.0245 | 7.70E-08 | **11000** | 45 × 45 | **6.43E+08** |
| L | Suspended | SiN | 3.20E-10 | 1.70E-02 | 5.44E-12 | 9.5 | 9.25~9.75 | 0.0729 | 1.10E-07 | 343 | 1000×1000 | 3.13E+08 |
| M | Suspended | AlN | 2.10E-09 | 4.40E-04 | 9.24E-13 | 8.8 | | 161 | 9.10E-09 | 4.31 | 200 × 75 | 5.80E+06 |
| N | Suspended | AlN | 1.90E-09 | 5.30E-03 | 1.01E-11 | 5 | | 172 | 5.80E-09 | 3.06 | 200 × 75 | 6.42E+06 |
| O | Suspended | GaN | 4.46E-07 | 5.56E-04 | 2.48E-10 | 0.88 | 0.76~1.1 | 101 | 7.50E-09 | 0.0168 | 80 × 80 | 17,937 |
| P | Suspended | AlN | 6.33E-10 | 3.50E-03 | 2.22E-12 | 5 | | 179 | 3.69E-09 | 7.128 | 60 × 144 | 1.45E+07 |
| Q | Suspended | SiN | 2.70E-11 | 5.00E-04 | 1.35E-14 | 8 | | 1.12 | 4.24E-08 | 1570 | 6 × 11 | 3.01E+07 |
| R | Suspended | AlN | 4.85E-10 | **1.66E-04** | 8.05E-14 | 4.5 | 2.75~6.25 | 825 | 2.67E-08 | 55 | 28 × 30 | 5.98E+06 |
| S | Suspended | AlN | 4.70E-08 | 5.30E-04 | 2.49E-11 | 5 | | 310 | 3.54E-09 | 0.76 | 75 × 200 | 2.54E+05 |
| T | Suspended | Quartz | 3.90E-09 | 5.92E-03 | 2.31E-11 | 10 | | 89.1 | N/A | N/A | N/A | 1.00E+05 |
| U | Suspended | AlN | 2.40E-09 | 1.40E-03 | 3.36E-12 | 4.5 | | 195 | 1.42E-09 | 1.68 | 100 × 200 | 5.89E+06 |
| V | Suspended | AlN | 3.40E-10 | 8.00E-04 | 2.72E-13 | 2 | 1.25~3.75 | 2,300 | N/A | N/A | 48 × 48 | 1.41E+07 |

(1) Due to the data availability, the definitions of NEP in different detectors might differ.

## Supplemental Notes

### 1. Design and FDTD simulation of absorbers

Our 6-μm single-wavelength absorber uses a C-shape design, while our 6/9-μm dual-wavelength absorber uses a double C-shape design. The dimensions of the absorbers are shown in **Fig. S7(a)**. We perform three-dimensional (3D) finite-difference time-domain (FDTD) method simulations using Ansys Optics (formerly, Lumerical) FDTD Solutions. **Figure S7(b)** shows the configuration of the FDTD simulation. The material optical properties of the LN were determined from Fourier transform infrared (FTIR) measurements of reflection spectra of the bulk LN substrate and LN thin films. We simulated one unit cell of absorber antennas on LN PnC structure. We use periodic boundary conditions in both *x* and *y* directions, and perfectly matching layers (PML) in both +*z* and -*z* directions. The period in the *x* direction is 2 μm, and the period in the *y* direction matches the antenna design. The dimension of the simulation region in the *z* direction is 1.8 μm. A plane wave with normal incident to the antenna is used as source. We set mesh option to 5 – "High accuracy", and the mesh size is about 6 nm at antenna region. We confirmed that all the simulations converged at an auto-shutoff value of $2 \times 10^{-5}$. Multiple field and power monitors are set at different heights extending the whole unit cell. Infrared absorption is derived from the net transmission between Monitors 1 and 3. Monitor 2 is used to show the optical field profiles in **Figs. 2(a) and 2(b)**.

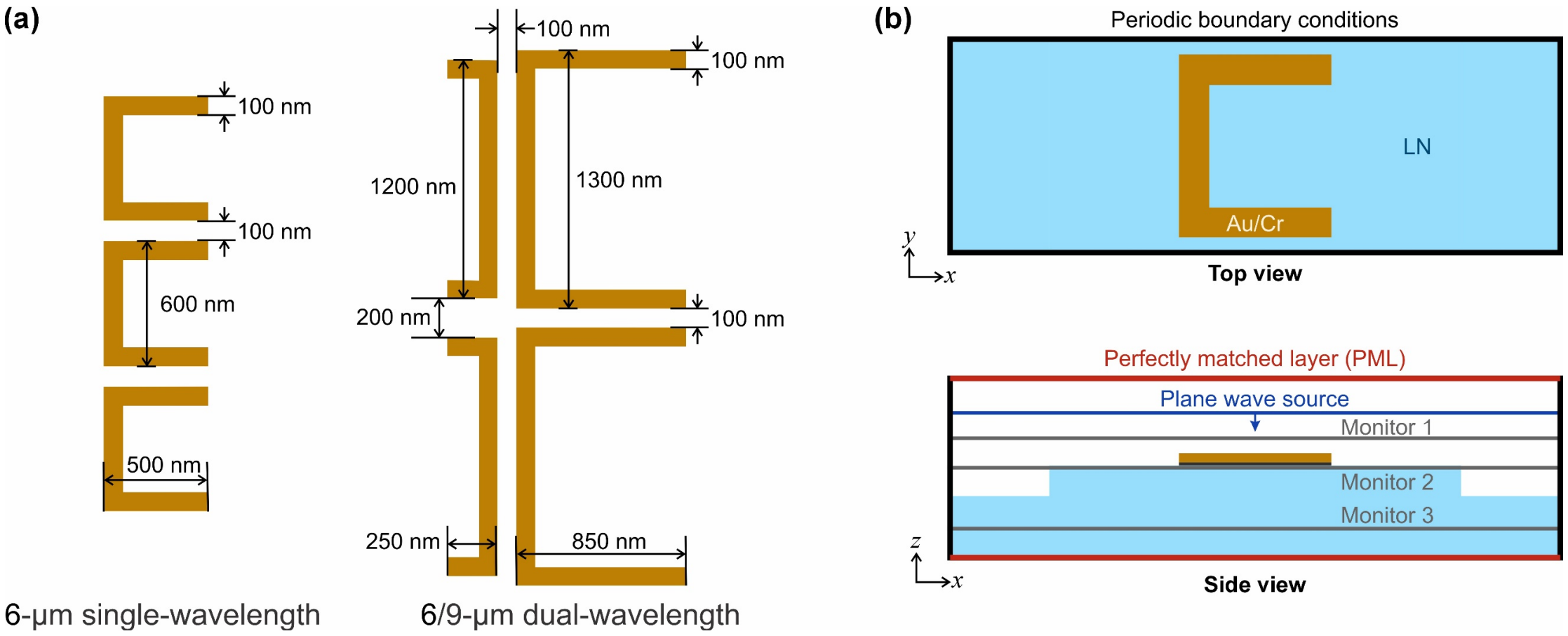

**Fig. S7 | Design and FDTD simulation of absorbers.** (a) Dimensions of the absorbers used in this work. (b) Configuration of the FDTD simulation.

### 2. Estimation of the acoustic mode area

We define the mode area $A_{\mathrm{mode}}$ of the acoustic mode by

$$A_{mode} = \frac{1}{u_{\mathrm{Max}}^2} \iint |u(x,y)|^2 \, dxdy,$$

where $u(x,y)$ is the measured displacement at local $(x,y)$ and $u_{\mathrm{MAX}}$ is the maximum displacement.

The acoustic mode area of the detector is calculated based on the measured mode profile in **Fig. 3(b)**, and we derive a mode area of 332 $\mu m^2$.

### 3. Estimation of the beam spot size and calibrated incident power

We use the knife-edge method to determine the 6-μm-wavelength beam spot size at the detector location. Gold patterns fabricated on quartz substrate are used as reflector, and scanned by a translational stage. The reflected power is monitored by a photodetector. We extract the 6-μm-wavelength beam diameters of 23 μm in x direction and 28 μm in y direction (**Fig. S8**). The measured result is consistent with the specification of the lens, which spec a beam diameter of 23 μm at 6 μm wavelength. Due to the experimental limitation, we use the beam diameter of 34 μm in the specification for the 9-μm-wavelength beam spot size. The calibrated incident power is determined by the integration of the input beam over the area with absorbers. The calibrated incident power over total beam power is 0.7073 for the 6-μm-wavelength beam and 0.6506 for the 9-μm-wavelength beam. The absorption area used for the calculation is 48 × 56 $\mu m^2$, with a mid-IDT width of 7.5 μm. The beam positioning is as follows: for the 6-μm-wavelength beam, it is 6 µm lower in the y-direction; for the 9-µm-wavelength beam, it is centered at the middle of the mid-IDT. Such calibration enables direct comparison with other works using different input optics and beam sizes.

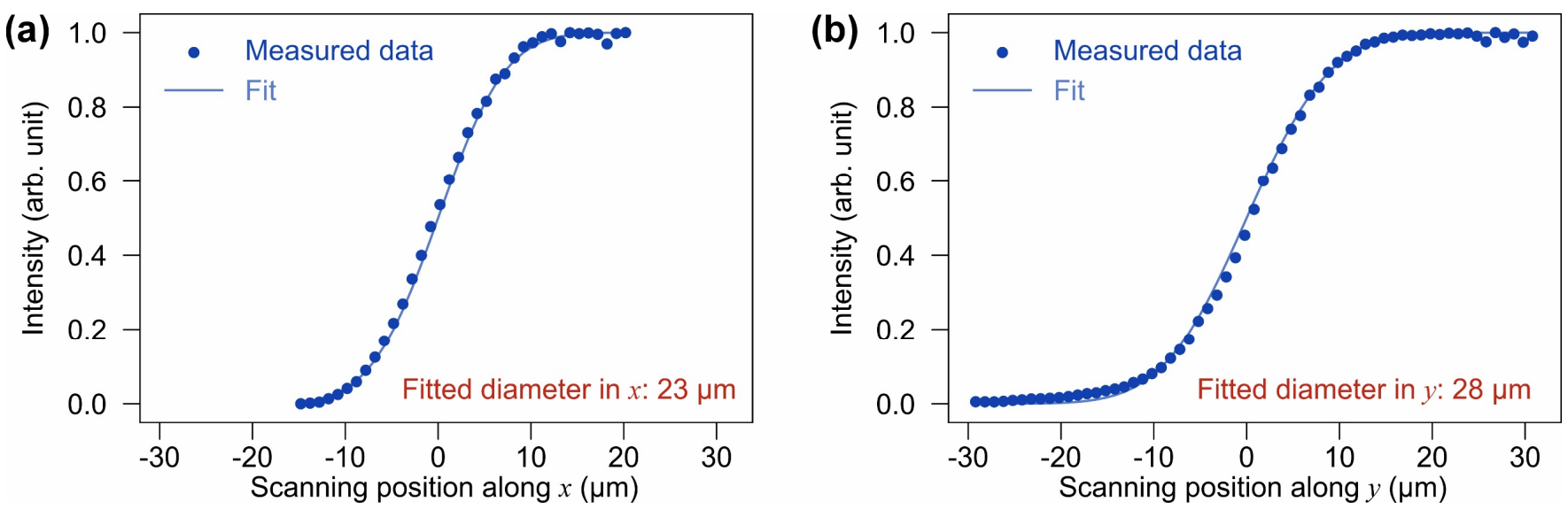

**Fig. S8 |** Measurement of the 6-µm-wavelength beam size using the knife-edge method. (a) The reflected intensity in the x direction scan. (b) Reflected intensity in the y direction scan.

## 4. Model of device dynamics

Our sensor involves (1) thermodynamics concerning the mid-infrared (MIR) input and thermal fluctuation noise and (2) dynamics of the mechanical oscillation. The thermodynamics is coupled parametrically to the mechanical dynamics through the temperature coefficient of the mechanical resonator.

The equation of motion of our sensor can be written as,

$$\frac{d}{dt}\Delta T(t) = -\frac{\Delta T(t)}{\tau_e} + \frac{P_{in}(t)}{c_P} + \zeta(t), \tag{S1}$$

$$\frac{d}{dt}a(t) = -i\left(\omega_0 + \alpha_T\omega_0\Delta T(t) + i\frac{g_{\mathrm{NL}}\big(a(t)\big)}{2} - i\frac{\gamma}{2}\right)a(t) + \sqrt{\kappa}\xi_{in}(t), \tag{S2}$$

where $\Delta T(t)$ is the temperature difference between the mechanical mode and the environment, $\tau_e$ is the thermal relaxation time to the environment, $P_{in}(t)$ is the absorbed power of the input MIR light, $c_p$ is the effective heat capacity of the mechanical mode, $\zeta(t)$ is the temperature fluctuation noise. As room temperature is much larger than the mechanical resonator frequency ($k_B T_0 \gg \hbar\omega$), $\zeta(t)$ can be described by a white Gaussian noise,

$$\overline{\zeta(t)} = 0, \qquad \overline{\zeta(t)\zeta(t')} = \frac{2\Delta T_0^2}{\tau_e}\delta(t-t'), \tag{S3}$$

where $\Delta T_0 = \sqrt{\frac{k_B T_0^2}{c_p}}$ is the total temperature fluctuation, $k_B$ is the Boltzmann constant, $T_0 = 300$ K is the temperature of the environment, $\delta(t-t')$ is the Dirac delta function. $a(t)$ is the amplitude of the mechanical resonant mode, in the unit of square root of phonon number, *i.e.*, $|a(t)|^2$ is the phonon number occupying the mechanical mode. $\omega_0$ is the natural resonant frequency of the mechanical mode. $\alpha_T$ is the normalized temperature coefficient. $\gamma$ is the total loss (including the intrinsic loss and the external coupling loss) of the mechanical mode, and the mechanical quality factor $Q = \omega_0/\gamma$. $\kappa$ is the external coupling rate of the mode. The saturable gain

$$g_{NL}(a) = \frac{g_0}{1 + \left|\frac{a}{a_{sat}}\right|^{\beta}}, \tag{S4}$$

where $g_0$ is the small signal gain, $a_{sat}$ is the saturation power, and the saturation parameter $\beta = 2$ matching the measured saturation curve of the amplified used in the work. When the gain fully compensates loss, $g_0 > \gamma$, the system will start self-oscillation and reach a stable state that $g = \gamma$. The input-referenced amplified thermal noise $\xi_{in}(t)$ can be described as a white Gaussian noise,

$$\overline{\xi_{in}(t)} = 0, \qquad \overline{\xi_{in}(t)\xi_{in}(t')} = A_{\mathrm{LNA}}^2\frac{4\,k_B T}{\hbar\omega_0}\delta(t-t'), \tag{S5}$$

where $A_{\mathrm{LNA}}$ is the noise amplification factor (sum of the gain and noise figure) of the low noise amplifier (LNA). We note that the thermal noise is amplified by the LNA. The output of the oscillator can be written as $a_{out}(t) = \sqrt{\kappa}\,a(t)$.

The sensing protocol is to estimate the input power $P_{in}(t)$ by observing the oscillator output $a_{out}(t)$. The complex value of $a_{out}(t)$ is observed by using an IQ demodulator and a local oscillator with frequency near $\omega_0$ (**Fig. S3(a)**).

## 5. Calculation of responsivity and noise equivalent absorbed power

When the system is noiseless, *i.e.*, all noise terms are set to 0, a constant input $P_{in}(t) = P_{in}$ will result in the shift of the oscillation frequency

$$\Delta\omega = \omega - \omega_0 = \alpha_T \omega_0 \Delta T = \frac{\alpha_T \omega_0 \tau_e}{c_p} P_{in}.$$

The DC responsivity $R_{DC} = \alpha_T \omega_0 \tau_e / c_p$.

By solving Eq. S1, we can obtain the noise equivalent power (NEP) $P_{\mathrm{NEP},\zeta}$ due to thermal fluctuation noise $\zeta$ as

$$P_{\mathrm{NEP},\zeta}\sqrt{\Delta t} = \sqrt{\frac{2\, k_B T_0^2\, c_p}{\tau_e}}.$$

We note that the left-hand side $P_{neq,\zeta}\sqrt{\Delta t}$ gives the normalized NEP in unit of W$\sqrt{\mathrm{s}}$, i.e., W/$\sqrt{\mathrm{Hz}}$.

For short term, we can estimate the oscillation frequency by the phase change, $\omega(t) = -\frac{d\phi}{dt}$. Note that $e^{-i\phi} \approx 1 - i\phi$, and $a(t + \Delta t) = a(t)e^{-i\phi}$. Solving Eq. *S*(2), the frequency noise $\Delta\omega = \overline{|\omega(t) - \omega_0|}$ due to the thermomechanical noise $\xi_{in}$ is given by

$$\Delta\omega\sqrt{\Delta t} = \frac{\sqrt{\kappa}\cdot\overline{|\xi_{in}(t)|}}{\overline{|a(t)|}} = \gamma\, A_{\mathrm{LNA}}\sqrt{\frac{k_B T}{p_{sat}}}\ .$$

where $p_{sat}$ is the saturation power of the low noise amplifier. The noise equivalent power $P_{\mathrm{NEP},\xi_{in}}$ due to the thermal mechanical noise $\xi_{in}$ is

$$P_{\mathrm{NEP},\xi}\sqrt{\Delta t} = \frac{\Delta\omega}{R}\sqrt{\Delta t} = \gamma\, A_{\mathrm{LNA}}\sqrt{\frac{k_B T}{p_{sat}}}\cdot\frac{c_p}{\alpha_T \omega_0 \tau_e}\ .$$

With the extract device parameters from experiments and numerical simulations (**Table S2**), we can estimate the absorbed-power NEP of 4.38 pW/$\sqrt{\mathrm{Hz}}$ due to temperature fluctuation and 0.62 pW/$\sqrt{\mathrm{Hz}}$ due to amplified thermomechanical noise. Therefore, the temperature fluctuation noise is the dominated noise of the current detector design. We note that the NEP calculated in this section is referring to the absorbed MIR power, i.e., assuming a perfect absorption and overlap with the mechanical mode of the input MIR light.

**Table S2. Parameters used in device model calculation and simulation**

| Parameter | Unit | Value |
|---|---|---|
| Specific heat capacity of LN, $c$ | J/(kg·K) | 628 |
| Density of LN, $\rho$ | kg/m$^3$ | 4,640 |
| Acoustic mode volume, $V_{mode}$ | μm$^3$ | 332 × 1.35 |
| $c_p$ | J/K | $1.305\times10^{-9}$ |
| $\tau_e$ | s | $1.69\times10^{-4}$ |
| $\alpha_T$ | K$^{-1}$ | $-71.2\times10^{-6}$ |
| $\omega_o/(2\pi)$ | Hz | $10^9$ |
| $R_{DC}/(2\pi)$ | Hz/W | $9.2\times10^9$ |
| $P_{\mathrm{NEP},\zeta}\sqrt{\Delta t}$ | pW/$\sqrt{\mathrm{Hz}}$ | **4.38** |
| $Q$ | | 1,000 |
| $\gamma/(2\pi)$ | Hz | $10^6$ |
| $A_{\mathrm{LNA}}$ | 1 | 22.387 |
| $p_{sat}$ | W | 0.063 |
| $\Delta\omega\sqrt{\Delta t}/(2\pi)$ | Hz/$\sqrt{\mathrm{Hz}}$ | $5.74\times10^{-3}$ |
| $P_{\mathrm{NEP},\xi}\sqrt{\Delta t}$ | pW/$\sqrt{\mathrm{Hz}}$ | **0.62** |

## 6. Numerical simulation of the detector dynamics

We numerically simulate detector dynamics using the equations of motion in Eqs. S1 and S2. We simulate the system in the rotating reference frame of $\omega_0$, a time step $dt = 2 \times 10^{-8}$ s and a total simulation time of 2 s. Simulations are running in physical units, no pre-scaling is performed. **Figure S9** show the response of our oscillator-based detector to a 1 nW absorbed power with on-off frequency at 190 Hz. The simulated results agree with our experiments.

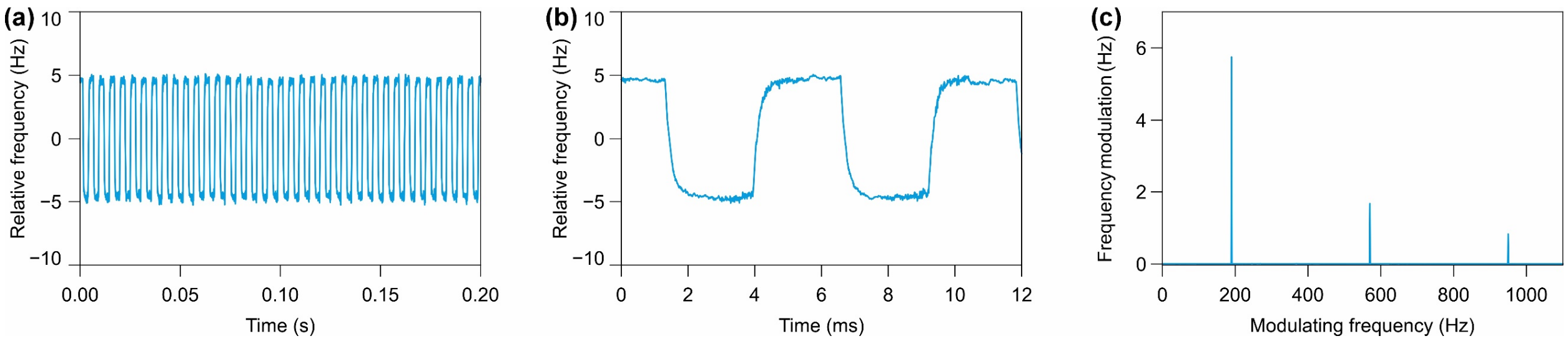

**Fig. S9 | Numerically simulated detector dynamics.** (a) Simulated oscillator frequency in response to a 1-nW absorbed power modulated by a 190 Hz square wave. (b) Zoom in view of (a). (c) Frequency spectrum of the frequency modulation signal, showing signals at the modulating frequency and its higher order harmonics.